\documentclass[pra,aps,showpacs,superscriptaddress,twocolumn]{revtex4}
\usepackage{color} 
\usepackage{amssymb} 
\usepackage{graphicx}
\usepackage[english]{babel}
\usepackage{epstopdf}
\usepackage{mathrsfs}
\usepackage{mathtools}
\usepackage{textcomp}
\usepackage{amsfonts}
\usepackage{amssymb} 
\usepackage{hyperref}

\newcommand{\mean}[2]{\left\langle#1\right\rangle_{#2}}

\newcommand{\op}[3]{\hat{#1}_{#2}^{#3}}

\newcommand{\be}{\begin{equation}}
\newcommand{\ee}{\end{equation}}
\newcommand{\bea}{\begin{eqnarray}}
\newcommand{\eea}{\end{eqnarray}}
\newcommand{\ba}{\begin{array}}
\newcommand{\ea}{\end{array}}

\begin{document}

\title{Spreading of correlations and Loschmidt echo after quantum quenches of a Bose gas in the Aubry-Andr\'e potential}
\author{Nicola Lo Gullo}
\author{Luca Dell'Anna}
\affiliation{Dipartimento di Fisica e Astronomia "G. Galilei" and CNISM, Universit\`a di Padova, 35131 Padova, Italy}

\begin{abstract}
We study the spreading of density-density 
correlations and the Loschmidt echo, after different sudden quenches in an 
interacting one dimensional Bose gas on a lattice, also in the presence of a 
superimposed aperiodic potential. We use a time dependent Bogoliubov 
approach to calculate the evolution of the correlation functions and employ 
the linked cluster expansion to derive the Loschmidt echo. 
\end{abstract}


\maketitle

\section{Introduction}
\label{sec:int}

The study of the behavior of many-body quantum systems driven out-of-equilibrium 
has attracted a lot of attention in the last few years.
In particular theoretical and experimental interest 
on how fast the correlations can spread in quantum many-body systems \cite{kollath2008,cheneau2012,ronzheimer2013,vidmar2013,carleo,jurcevic2014} has been 
renewed after the work by Calabrese and Cardy \cite{calabrese2006}. 
They showed that, for critical theories the maximum velocity of the 
spreading of 
correlations is given by the group velocity in the final gapless system. 
Actually the existence of a maximal velocity \cite{lieb1972}
known as the Lieb-Robinson bound, 
has been shown to exist theoretically in several interacting many-body systems, 
due to short range interactions which may reduce 
the propagation of information making its spreading speed finite.

In this work we study the spreading of 
density-density correlations
following sudden quantum quenches in a system of 
bosons held in a bichromatic lattice.
The case of bosons placed on a lattice is a paradigm of interacting 
many-body systems, which can be experimentally reproduced 
by means of ultracold atomic gases and described 
theoretically by the well known Bose-Hubbard model.
Beyond the maximum velocity, one can wonder how correlations evolve at later 
times. It has been shown \cite{natu2013} that, for bosons on a periodic 
lattice, density-density correlations spread diffusively after an initial 
ballistic motion. 
One issue worth being addressed is therefore related to the effects of a 
modulated potential on such behaviors. 
We were inspired by a recent experimental work \cite{modugno2013} 
in which transport of bosons in a bichromatic optical lattice 
was studied.

Besides the correlation spreading there are other quantities, 
useful to characterize the dynamics of a quantum system
and its approach to equilibrium, if any.
The Loschmidt echo is perhaps one of the most 
used tools to investigate the dynamics of a quantum
system following a sudden quench.
Physically, it is the probability for 
the system to return to its initial state after a certain time.
It is particularly sensitive to both the initial state
and the spectrum of the system after the sudden quench
and it thus reveals critical behaviors of the system \cite{sindona2013,schiro}.
Moreover it has been shown that the echo is 
related to the work distribution, which in turn is a very useful quantity when 
thermodynamical properties of a 
closed quantum system are considered \cite{sindona2014}.

In this paper we show how to calculate 
the evolution of correlation functions 
in interacting bosonic systems 
by means of a time dependent Bogoliubov approach \cite{mora2003}, 
and derive non-perturbatively the Loschmidt echo, 
by means of linked cluster expansion \cite{mattuck}.

\section{Model and method}
\label{sec:meth}
We consider a system of interacting bosons in a 1D lattice 
with on-site interaction. 
In the single band approximation, this system is described by 
the the Bose-Hubbard Hamiltonian

\begin{equation}
\label{eq:totham}
{\hat {\cal H}}=-\frac{J}{2}\sum\limits_{\langle i,j\rangle}^L \hat{b}_i^\dag \hat{b}_{j}+\sum\limits_{i}^L V_{i}\,\hat{b}_i^\dag \hat{b}_i
+\frac{U}{2}\sum\limits_{i}^{L} \hat n_i(\hat n_i-1),
\end{equation}

\noindent where $\hat b^\dag_i$ and $\hat b_i$ are bosonic creation 
and annihilation operators defined on the lattice sites, 
$V_i$ are the on-site energies, 
$J$ the hopping parameter between nearest neighbor sites, 
$U$ the on-site boson-boson interaction, $\hat n_i=\hat b^\dag_i \hat b_i$ 
the number operator and $L$ the number of sites.
In what follows we will consider a modulation of the
on-site potential of the Aubry-Andr\'e type (also known as Harper model),

\begin{equation}
\label{VAA}
 V_{i}=\lambda \cos\left(2\pi \tau\,i\right),
\end{equation}

\noindent 
where we choose $\tau=(\sqrt{5}+1)/2$, the golden ratio.
In the non-interacting case, $U=0$, 
it has been proven rigorously \cite{jitomirskaya1999} that
the above system shows a metal-insulator like transition 
at $\lambda=\lambda_c=1$ {(here and in what follows we assume $J=1$)}. 
For $\lambda>\lambda_c$ all eigenstates are exponentially localized, 
while in the case $\lambda<\lambda_c$ are all delocalised.
This peculiarity leads, in the presence of weak interaction, 
to the existence of a superfluid state even for finite values of $\lambda$, 
in contrast to uncorrelated disorder, which, even in the presence of a 
small amount, 
is more effective to bring the system to a Bose glass phase.
{
This behavior has been confirmed in several works where the phase diagram
of the model described by Eq. (\ref{eq:totham})
has been derived \cite{roux2008,roscilde2008,deng2008}, 
showing that, for $\lambda<1$ and moderate interaction, 
the system is in a superfluid phase.}

This allow us to safely address the case of weakly interacting bosons 
at zero temperature by means of the time-dependent Bogoliubov 
approach even at finite values of $\lambda$.
{In the high filling limit, 
weak boson-boson interaction plays an important r\^ole,
not because of particle interaction but because of the eventually large number 
of  particles on single sites \cite{dutta2011}. 
We assume, therefore, $U\mean{\op{n}{i}{}}{}$ not too large, and consider small quantum fluctuations.} 

{In this limit we can separate the bosonic operator into a spatially varing 
classical part} (mean field) and a quantum part (quantum fluctuations)

\be
\label{b}
\hat{b}_i=\sqrt{N_0}\,\phi_i+\hat{c}_i
\ee
with $N_0$ the (macroscopic) number of particles occupying the state
$\phi_i$.

Within gaussian approximation in the fluctations, 
we get the following effective Bogoliubov Hamiltonian 

\bea
\label{eq:Hb}
\hat H&=&\sum_{i}\left(V_i-\mu+2g |\phi_i|^2\right)\hat c^\dag_i \hat c_i-\frac{J}{2}\sum_{\langle i,j\rangle}\hat c^\dag_i \hat c_{j}\\
&&+\frac{g}{2}\sum_i (\phi_i^2\,\hat c_i^\dag \hat c_i^\dag+\phi^{*2}_i\,\hat c_i \hat c_i )\nonumber,
\eea
where $g=U N_0$. 
The macroscopically occupied state $\phi_i$ and the chemical potential
$\mu$ satisfy the stationary Gross-Pitaevskii equation 
\be
\label{eq:GP}
-\frac{J}{2}(\phi_{i+1}+\phi_{i-1})+g|\phi_i|^2\phi_i+V_i\phi_i=\mu\phi_i .
\ee

The Hamiltonian in Eq.(\ref{eq:Hb}) can be diagonalized by means of the 
Bogoliubov transformations (see Sec. \ref{sec:ququench}).
We solved Eq. (\ref{eq:GP}) and diagonalized Eq.(\ref{eq:Hb})
iteratively by fixing the total number of particles in a system 
of $L=100$ sites to be $N=500$, setting $N_0=N-N_{ex}$ where $N_{ex}$ 
is the number of particles in the excited states. 
Usually after five iterations the solution converges {and we checked 
that $N_{ex}\ll N_0$ for all ranges of parameters we have used, in 
accordance with the assumtion of small fluctuations.}

\begin{figure}[h]
 \includegraphics[width=8.3cm]{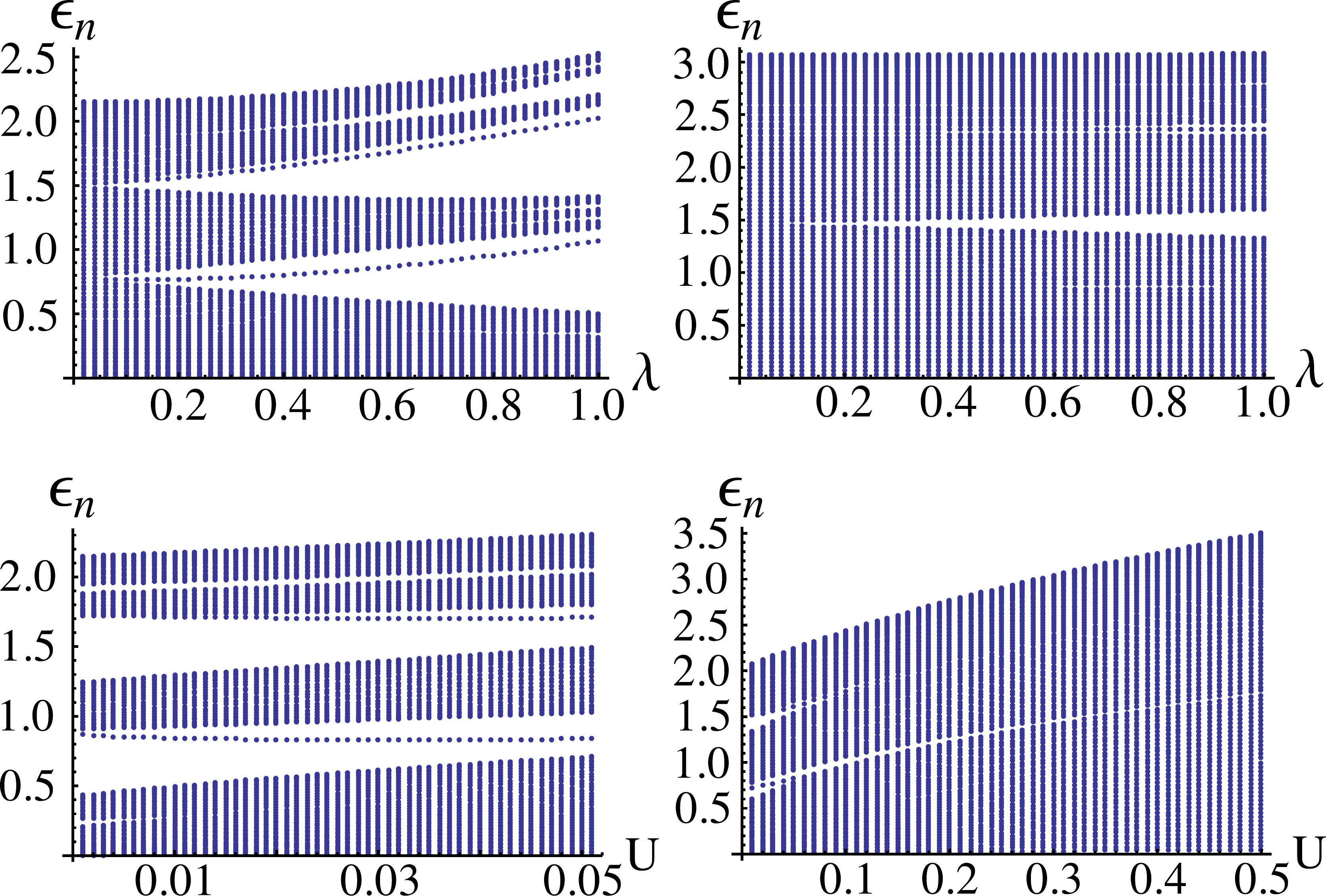}
\caption{Bogoliubov {bands} (top row) as functions of $\lambda$ at (left) 
$U=0.03$ and (right) $U=0.3$ 
and (bottom row) as functions of $U$, at $\lambda=0.5$, {for small (left) and large (right) values of $U$.}}
 \label{fig:spectr}
\end{figure}

{In Fig.\ref{fig:spectr} we plot the band spectrum of the Bogoliubov modes
as a function of $\lambda$ at fixed $U$, and as a function
of $U$ at fixed $\lambda$. 
One can notice that the effect of the interaction is not only that of closing
the sub-bands, as expected, but also making} the inter-band localized state 
to migrate from the higher energy sub-band to the lower energy one.

\section{Quantum quenches}
\label{sec:ququench}

In this section we present the formalism used to look at the dynamics 
of the system following a sudden quench in the Hamiltonian.
In the following sections we will consider quenches 
resulting from a sudden change in
i) $U$ at fixed $\lambda$,
ii) $\lambda$ at fixed $U$.

It is worth mentioning that with the Bogoliubov approach
the system is closed but not isolated.
In fact the system described by the effective Hamiltonian in Eq. (\ref{eq:Hb})
does not conserve the total number of quasi-particles if a parameter
is changed.
This is due to the fact that also the chemical potential changes
and the system can exchange particles with the superfluid part.
Therefore we are dealing with a system, which can exchange particles
with a reservoir.

Let us call $H_0$ the initial Hamiltonian at time $t=0$, 
described by Eq. (\ref{eq:Hb}) 
with initial parameters ($U=U_0$, $V_i=V_{0i}$, and, 
from Eq. (\ref{eq:GP}), $\mu=\mu_0$, {$N_0=N_{00}$}) while 
{$H$ in Eq. (\ref{eq:Hb}) is the Hamiltonian after the quench.} 
Both $H_0$ and $H$ can be diagonalized 
by the following canonical Bogoliubov transformations

\bea
\label{bog1}
\hat c_i&=&\sum_n u_{i,n} \,\hat \alpha_n -v^*_{i,n}\, \hat \alpha^\dag_n\,,\\
\hat c_i
&=&\sum_n \omega_{i,n} \,\hat \beta_n - w^*_{i,n}\,\hat \beta^\dag_n\,, 
\label{bog2}
\eea
%
with conditions $\sum_i(u_{in}u^*_{im}-v_{in}v^*_{im})=\sum_i(\omega_{in}\omega^*_{im}-w_{in}w^*_{im})=\delta_{nm}$ , ensuring that the above transformation are indeed canonical, 
so that by Eqs. (\ref{bog1}), (\ref{bog2}) we get
\bea
&&\hat H_0=\sum_n \epsilon^0_{n}\,\hat \alpha^\dag_n \hat\alpha_n\,,\\
&&\hat H=\sum_n \epsilon_{n}\,\hat \beta^\dag_n\hat\beta_n\,,
\eea
where $n$ labels the eigenmodes. 
Thus we can write the operators $\hat \beta_n$ of 
the diagonalized final Hamiltonian in terms of the 
Bogoliubov operators $\alpha_n$ of the initial Hamiltonian

\be
\label{beta}
\left(\ba{c}
\hat\beta_n\\
\hat\beta_n^\dag\ea\right)=\sum_m\left(\ba{cc}
\Lambda_{nm}&\Omega^*_{nm}\\
\Omega_{nm}&\Lambda^*_{nm}
\ea\right)\left(\ba{c}
\hat\alpha_m\\
\hat\alpha_m^\dag\ea\right)
\ee 
where
\bea
\Lambda_{nm}=\sum_i \omega^*_{i,n} u_{i,m}-w^*_{i,n} v_{i,m}\\
\Omega_{nm}=\sum_i w_{i,n}u_{i,m}-\omega_{i,n}v_{i,m}
\eea
When $H=H_0$, $\Lambda_{nm}=\delta_{nm}$ and $\Omega_{nm}=0$
due to the conditions on the coefficients of the transformations.
The initial state $|\psi(0)\rangle$ is chosen to be the 
vacuum state of $H_0$, namely $\alpha_n|\psi_0\rangle=0$ $\forall n$, 
while the evolution of the original operators in the Heisenberg picture, 
is given by

\be
\label{ct}
\hat c_i(t)=\sum_n \omega_{i,n} \,\hat \beta_ne^{-i \epsilon_n t} - w^*_{i,n}\,\hat \beta^\dag_n e^{i\epsilon_n t},
\ee
where $\epsilon_n$ are the Bogoliubov energies of the final Hamiltonian $H$. 
Thus we are able to calculate the time evolution of all correlation functions, 
within the gaussian approximation, on the {ground state of $H_0$}, 
by means of Eq. (\ref{beta}).

\section{Correlation functions}
\label{sec:corrfuncs}
In what follows we will consider the normal ordered density-density 
correlators between different sites at different times, 
\be
{\cal G}_{i,j}(t,t')=
\langle:{\hat n}_i(t) \hat n_j(t'):\rangle-\langle \hat n_i(t) \rangle
\langle \hat n_j(t')\rangle\,.
\ee
At the leading order in the fluctuations, neglecting variation of $\phi$ 
for small quenches and using Eq. (\ref{b}), we have
\be
\label{eq:G}
{\cal G}_{i,j}(t,t')\simeq 2N_0 \textrm{Re}
\left[\phi_i\phi_j^{*}\,\langle \hat c_i^\dag(t)\hat c_j(t')\rangle
+\phi^*_i\phi^*_j\,\langle \hat c_i(t)\hat c_j(t')\rangle\right].
\ee
Therefore, we need to calculate only $\langle \hat c_i^\dag(t)\hat c_j(t')\rangle$ and $\langle c_i(t)c_j(t')\rangle$.
From Eq. (\ref{ct}) and its conjugate counterpart, and Eq. (\ref{beta}), 
and exploiting the fact that the initial state is the vacuum state 
of the $\hat \alpha$'s, we get

\begin{widetext}
\bea
\nonumber
\langle \hat c_i^\dag(t) \hat c_j(t')\rangle=\sum_{n,\ell, m} \left\{
\omega^*_{i,n}\omega_{j,\ell}
\,\Omega_{nm}\Omega^*_{\ell m}\,e^{i(\epsilon_n t-\epsilon_\ell t')}
+w_{i,n} w^*_{j,\ell}
\,\Lambda_{nm}\Lambda^*_{\ell m}\,e^{-i(\epsilon_n t-\epsilon_\ell t')}\right.\\
\left. -\omega^*_{i,n}w^*_{j,\ell}
\,\Omega_{nm}\Lambda^*_{\ell m}\,e^{i(\epsilon_n t+\epsilon_\ell t')}
-w_{i,n}\omega_{j,\ell}
\,\Lambda_{nm}\Omega^*_{\ell m}\,e^{-i(\epsilon_n t+\epsilon_\ell t')}
\right\}\\
\nonumber \phantom{.}\\
\nonumber
\langle \hat c_i(t) \hat c_j(t')\rangle=\sum_{n,\ell, m} \left\{
\omega_{i,n}\omega_{j,\ell}
\,\Lambda_{nm}\Omega^*_{\ell m}\,e^{-i(\epsilon_n t+\epsilon_\ell t')}
+w^*_{i,n} w^*_{j,\ell}
\,\Omega_{nm}\Lambda^*_{\ell m}\,e^{i(\epsilon_n t+\epsilon_\ell t')}\right.\\
\left. -\omega_{i,n}w^*_{j,\ell}
\,\Lambda_{nm}\Lambda^*_{\ell m}\,e^{-i(\epsilon_n t-\epsilon_\ell t')}
-w^*_{i,n}\omega_{j,\ell}
\,\Omega_{nm}\Omega^*_{\ell m}\,e^{i(\epsilon_n t-\epsilon_\ell t')}
\right\}.
\eea
\end{widetext}
In what follows we will look at the following function

\be
\label{eq:DGtt}
\Delta{\cal G}_{i}(t,t')={\cal G}_{i,i_0}(t,t')-{\cal G}_{i,i_0}(0,0)
\ee
with $i_0$ a fixed point of the lattice, which in the following will
be chosen to be $i_0=L/2$. 
In particular we will look at $\Delta{\cal G}_{i}(t,0)$, which gives us 
information on 
the propagation of the effect of a perturbation acting at $i_0$ at time $t=0$
after some time $t$ at a point $i$.

{The density-density correlation function at different times, in 
Fourier space, is also called dynamical structure factor.
This quantity has been calculated for the Lieb-Liniger model 
\cite{calabrese2006_2}, namely for a 1D Bose gas in the continuum. 
As also reported in Ref. \cite{calabrese2006_2}, 
the dynamical structure factor 
and, therefore, the density-density correlation function at different times, is
experimentally accessible either by Fourier sampling of time of flight images 
\cite{duan2006} or through Bragg spectroscopy \cite{ketterle1999}. 
More recently a direct, real-time and nondestructive measurement of the 
dynamic structure factor has been realized for a Bose gas to reveal a 
structural phase transition \cite{esslinger}.}

\subsection{Periodic case  $(\lambda=0)$}
We start the discussion about the behavior of the density-density
correlation functions by first looking at the homogeneous case.
In this case only a quench in the boson-boson interaction $U$
can be performed.

\begin{figure}
    \includegraphics[width=4.1cm]{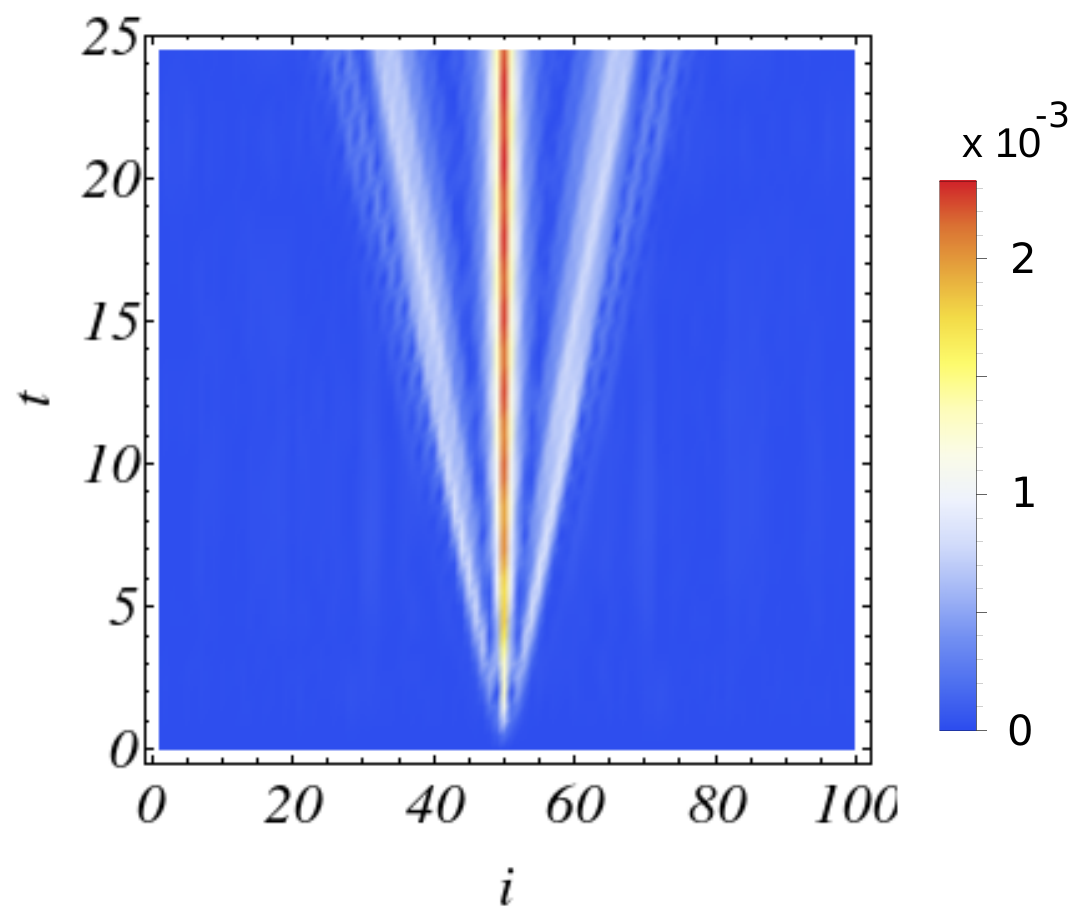}
    \includegraphics[width=4.1cm]{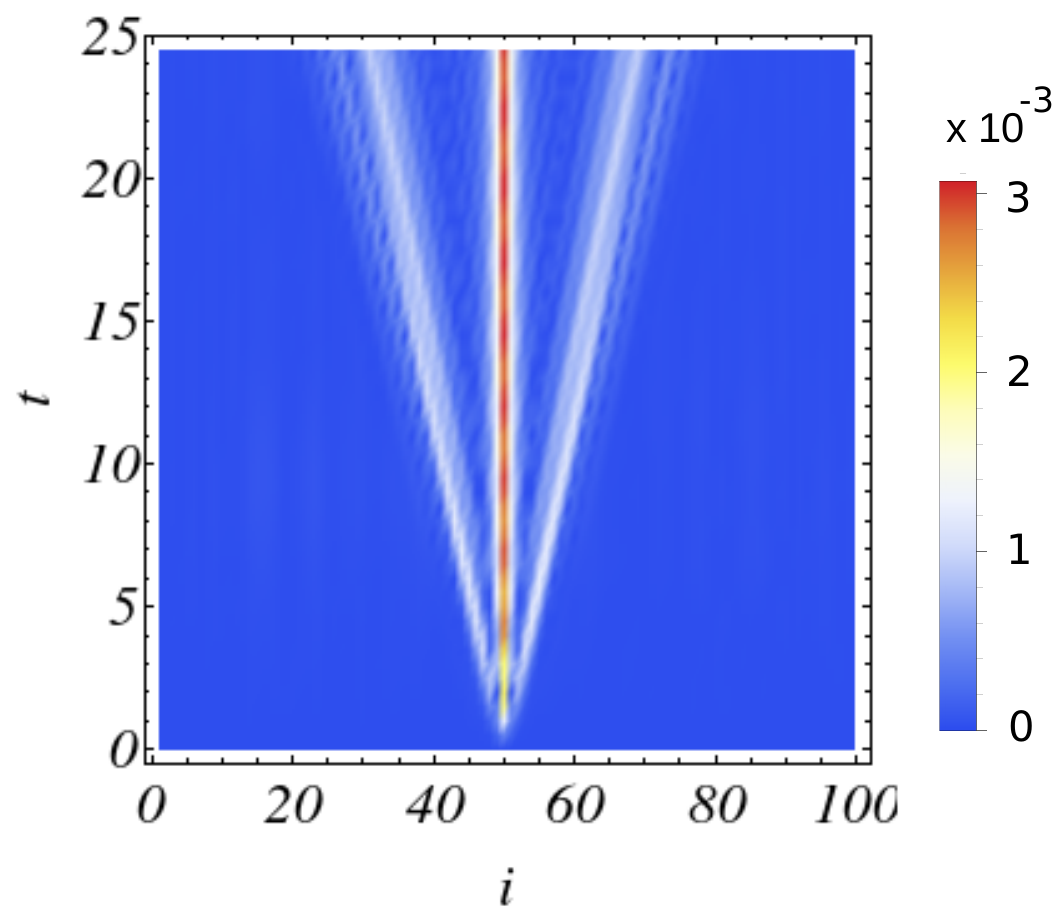}
    \includegraphics[width=4.1cm]{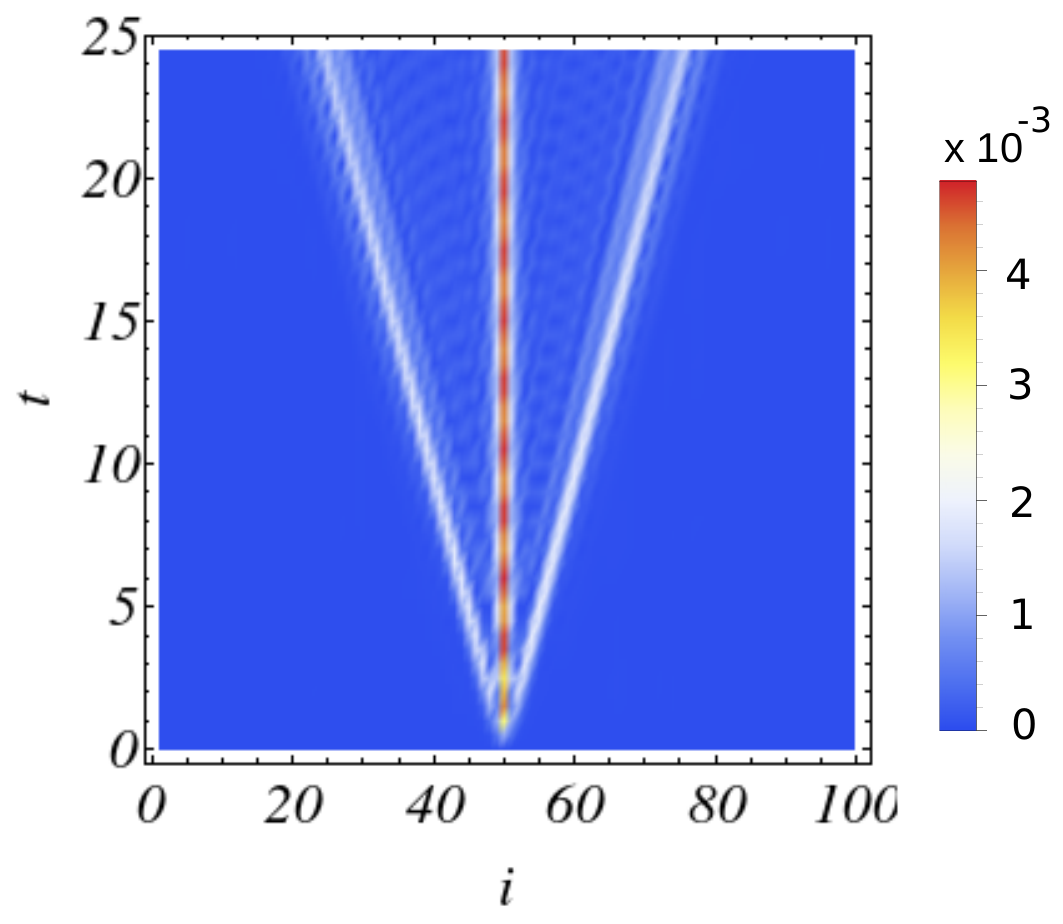}
    \includegraphics[width=4.1cm]{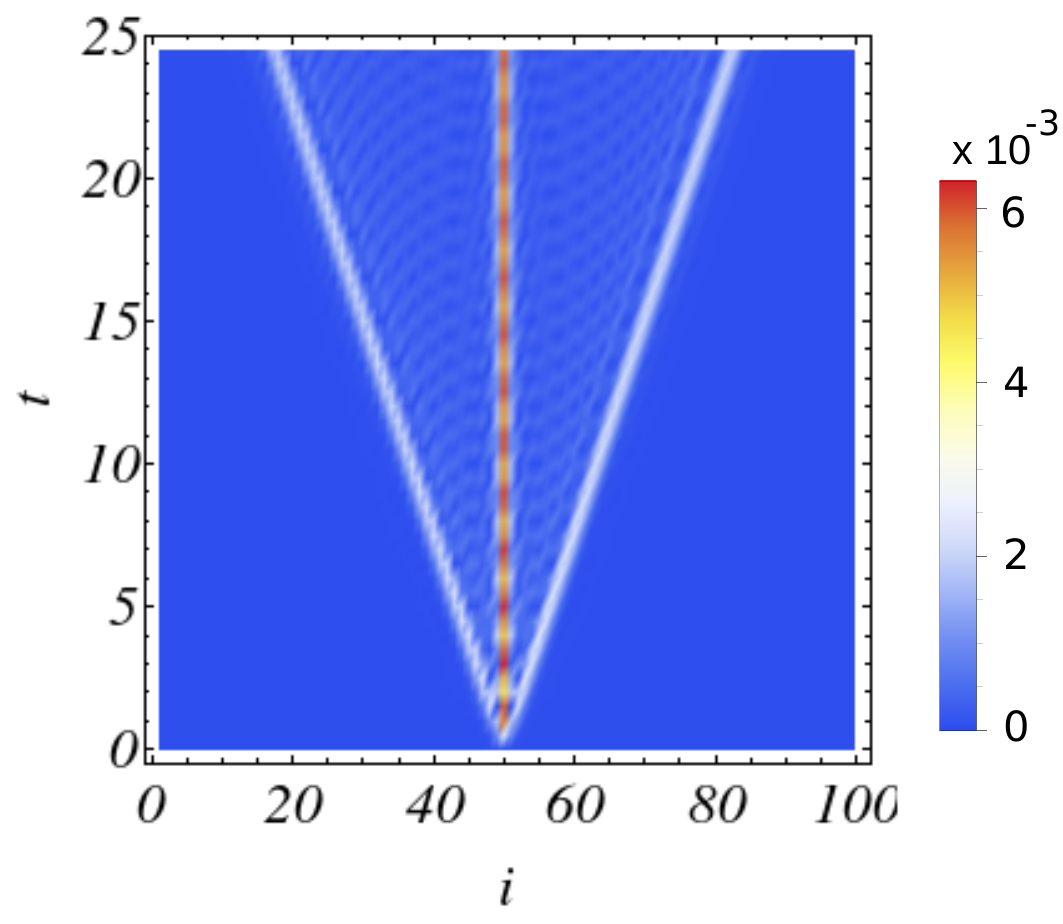}
\caption{(Color online) $\Delta{\cal G}_{i}(t,0)$, Eq. (\ref{eq:DGtt}), after 
quenches from $U_0=0.025$ (top left), $0.05$ (top right), $0.15$ (bottom left), $0.35$ (bottom right), to $U=U_0+0.05$, as a function 
of the distance $i$ and time $t$ at $\lambda=0$. {Here and in all the figures, $t$ is in unit of $J^{-1}$ and $i$ in unit of the lattice spacing.}}
 \label{fig:homquenchU}
\end{figure}

\begin{figure}
    \includegraphics[width=6.3cm,height=6.1cm]{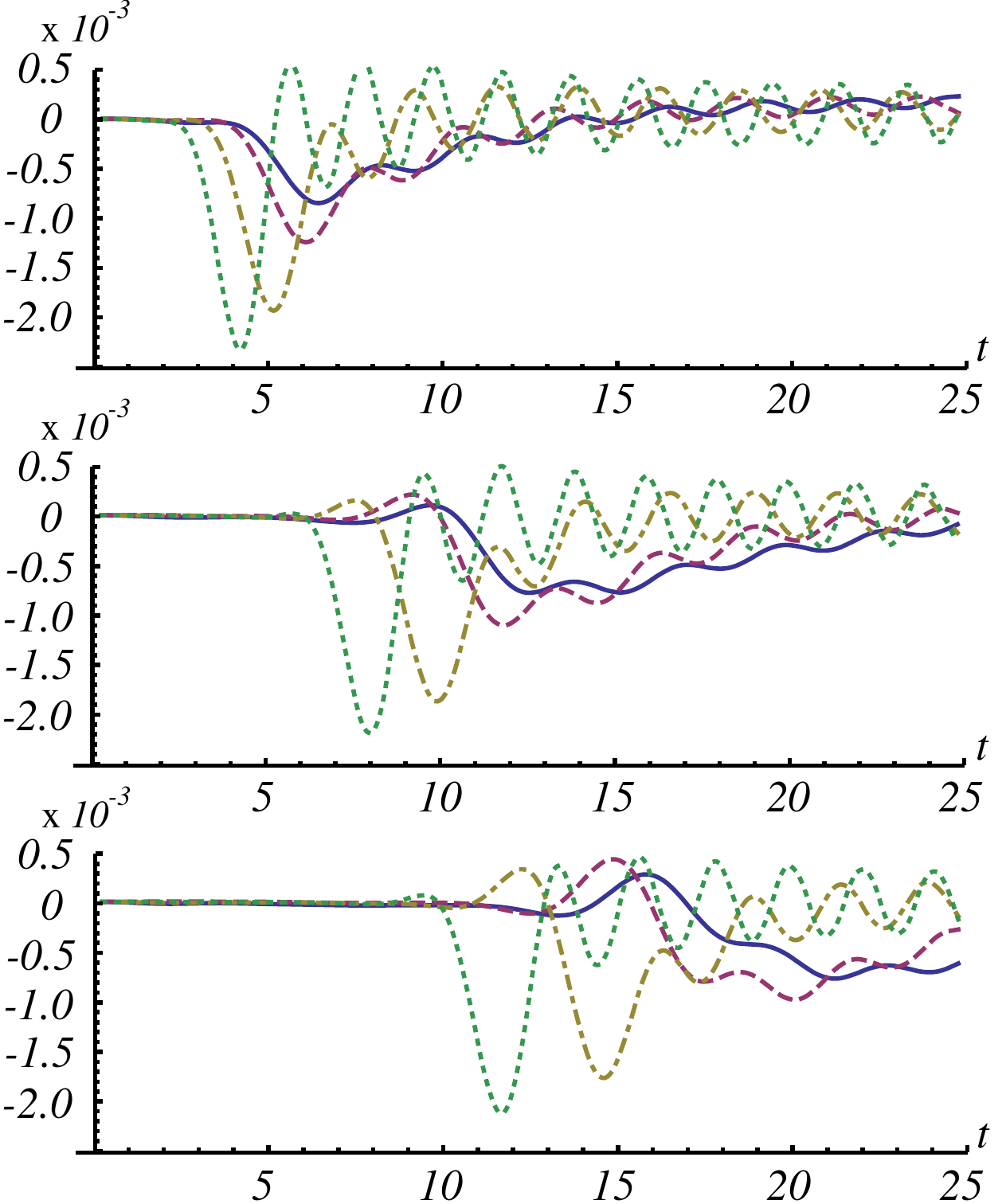}
\caption{(Color online) $\Delta{\cal G}_{i}(t,0)$ for $i-i_0=5,10,15,20$, 
after a quench from $U_0=0.025$ (solid line), $U_0=0.05$ (dashed line), 
$U_0=0.15$ (dot-dashed line), $U_0=0.35$ (dotted line), 
as a function of time $t$. $U-U_0= 0.05$ for all quenches.}
 \label{fig:detquenchU}
\end{figure}

{
In Fig. \ref{fig:homquenchU} we show the propagation
of density-density correlations.  
We can clearly see that the fastest signal is ballistic 
and that the speed of propagation increases by increasing $U_0$ together with its amplitude (see Fig. \ref{fig:detquenchU}), while the 
slower diffusive part is very intense for small $U_0$ and almost disappear 
for large $U_0$. \\
The velocity of the fast signals is, therefore, constant and given by the maximum value of the group velocity in the final system \cite{calabrese2006}, which, from the single particle dispersion $\varepsilon_k=J(1-\cos k)$ and the 
Bogoliubov spectrum $\epsilon_k=\sqrt{\varepsilon_k(\varepsilon_k+2\nu U)}$, 
is given by
\be
\label{speed}
v=\frac{J(\varepsilon_{k_m}+\nu U)\sin k_m}{\epsilon_{k_m}}
\ee 
where $\nu$ is the filling and 
\be
k_m=2\arccos\sqrt{\frac{3}{4}+\frac{3\nu U}{J}-\frac{\sqrt{J^2(J^2+6J\nu U+5\nu^2U^2)}}{4J^2}}.
\ee
For $U\simeq 0.4$, $\nu\simeq 5$ and $J=1$, the speed given by the Eq. (\ref{speed}) is $v\approx 1.4$, in agreement with the speed of the propagation shown in the last pannel of Fig. \ref{fig:homquenchU}, with same parameters. For relatively small $\nu U$, we can approximate $k_m\simeq \pi/2$ and, therefore, the speed is simply given by
\be
v\simeq \frac{J(J+\nu U)}{\sqrt{J(J+2\nu U)}}.
\ee
}

\subsection{Quench in $U$}
Let us first consider the quench in the boson-boson interaction $U$. 
For this {case} we can compare the results in the absence of a 
modulation of the on-site energies ({$\lambda=0$}) 
with those obtained at finite $\lambda$. 

\begin{figure}[h]
\includegraphics[width=4.1cm]{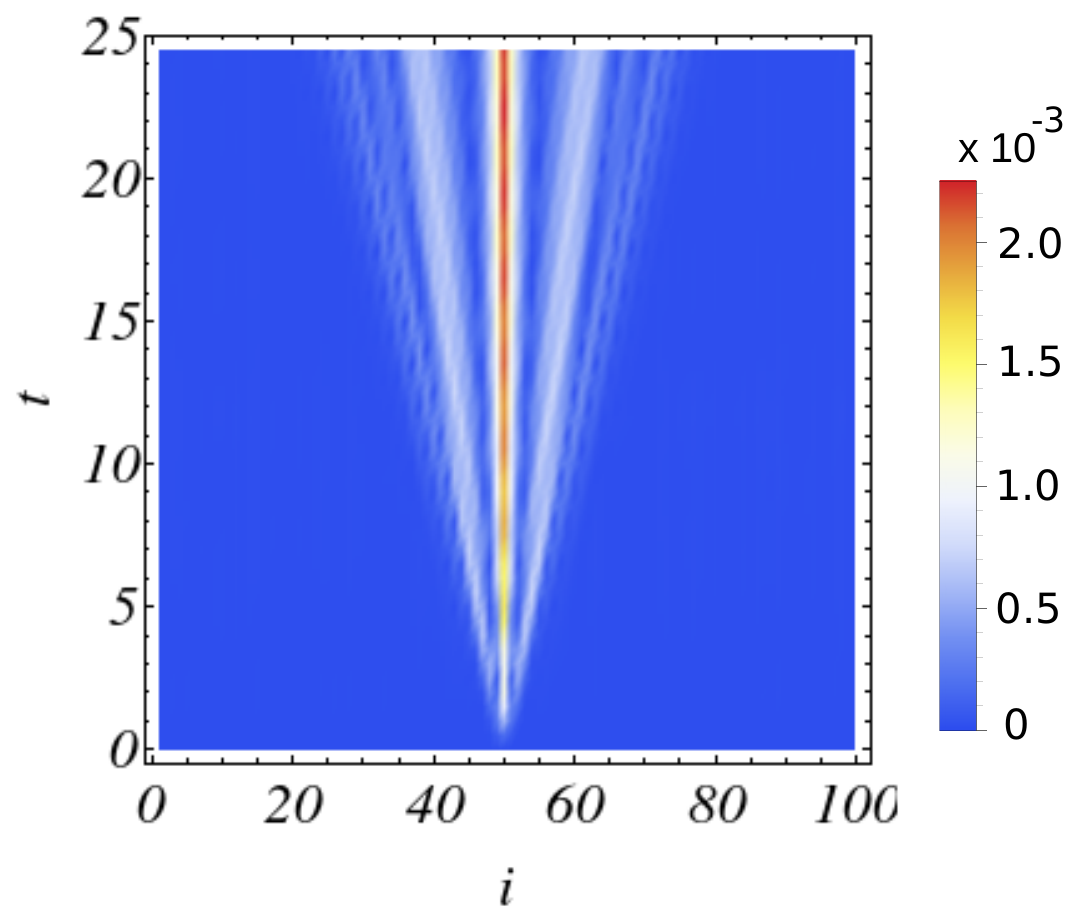}
\includegraphics[width=4.1cm]{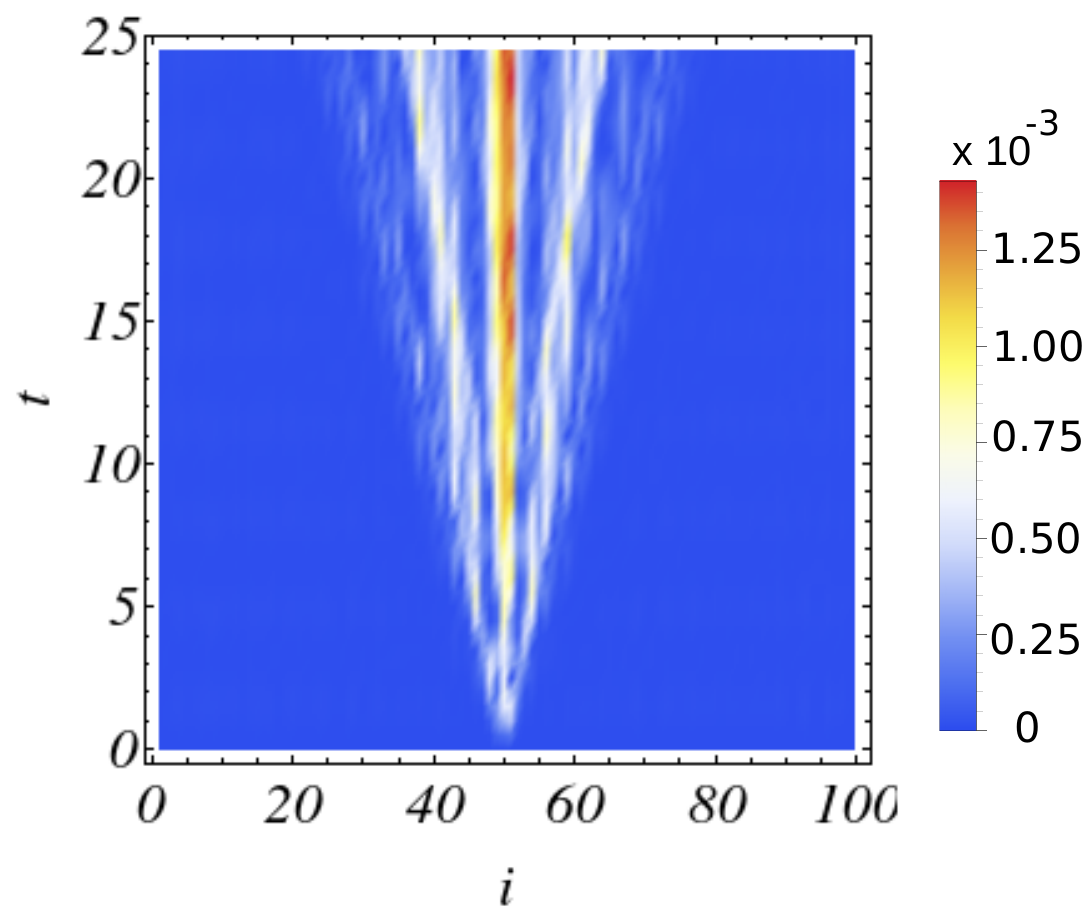}
\includegraphics[width=4.1cm]{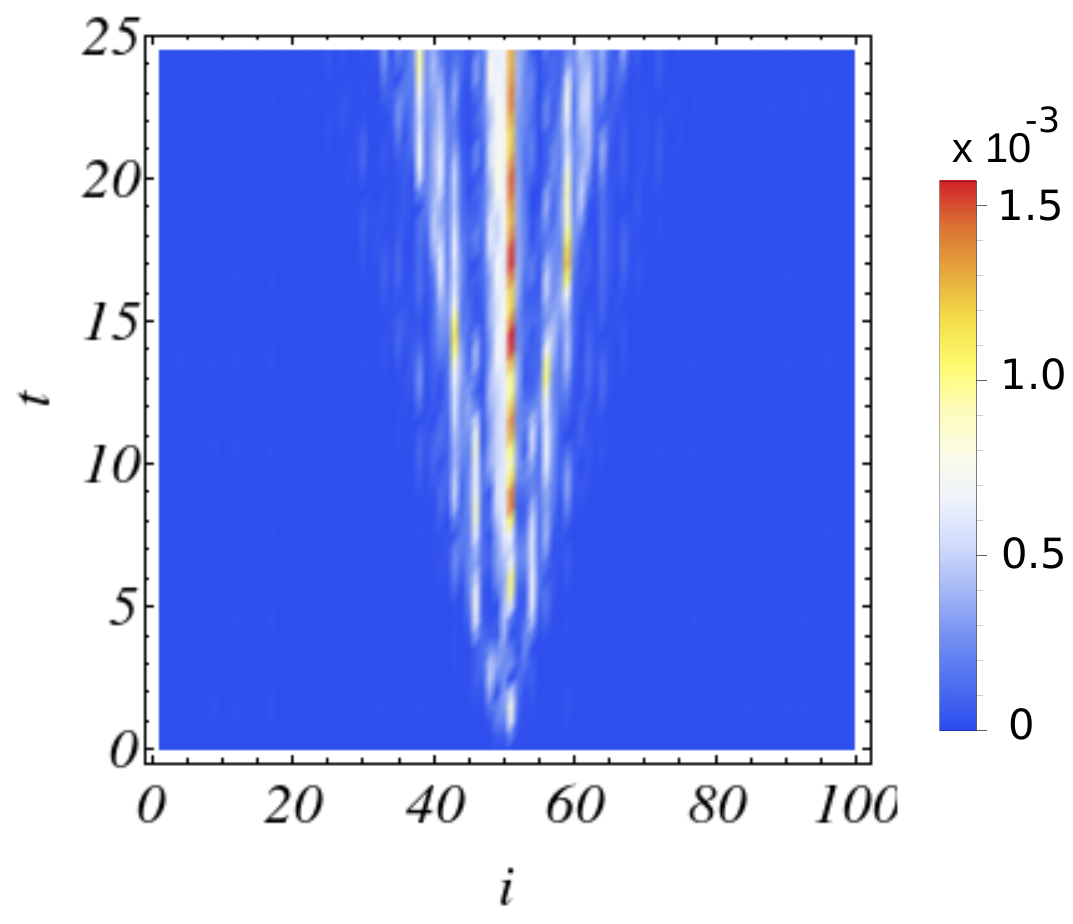}
\includegraphics[width=4.1cm]{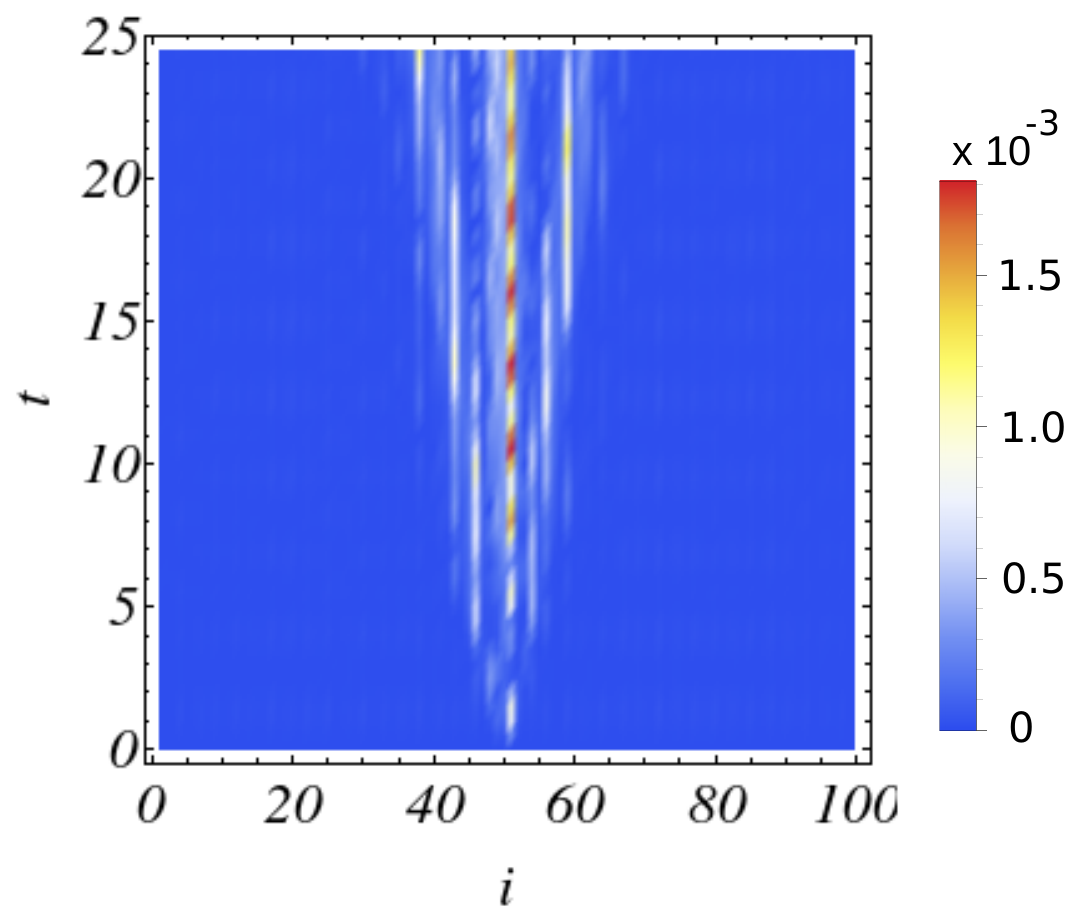}
\caption{(Color online) 
$\Delta{\cal G}_{i}(t,0)$, Eq. (\ref{eq:DGtt}), after a 
quench from $U_0=0.025$ to $U=0.03$, as a function of the 
distance $i$ and time $t$, {for different values of $\lambda$: 
$\lambda=0$ (top left), $0.3$ (top right), $0.6$ (bottom left), $0.9$ (bottom right).}}
 \label{fig:quenchU0.025}
\end{figure}

\begin{figure}[h]
    \includegraphics[width=6.3cm,height=6.1cm]{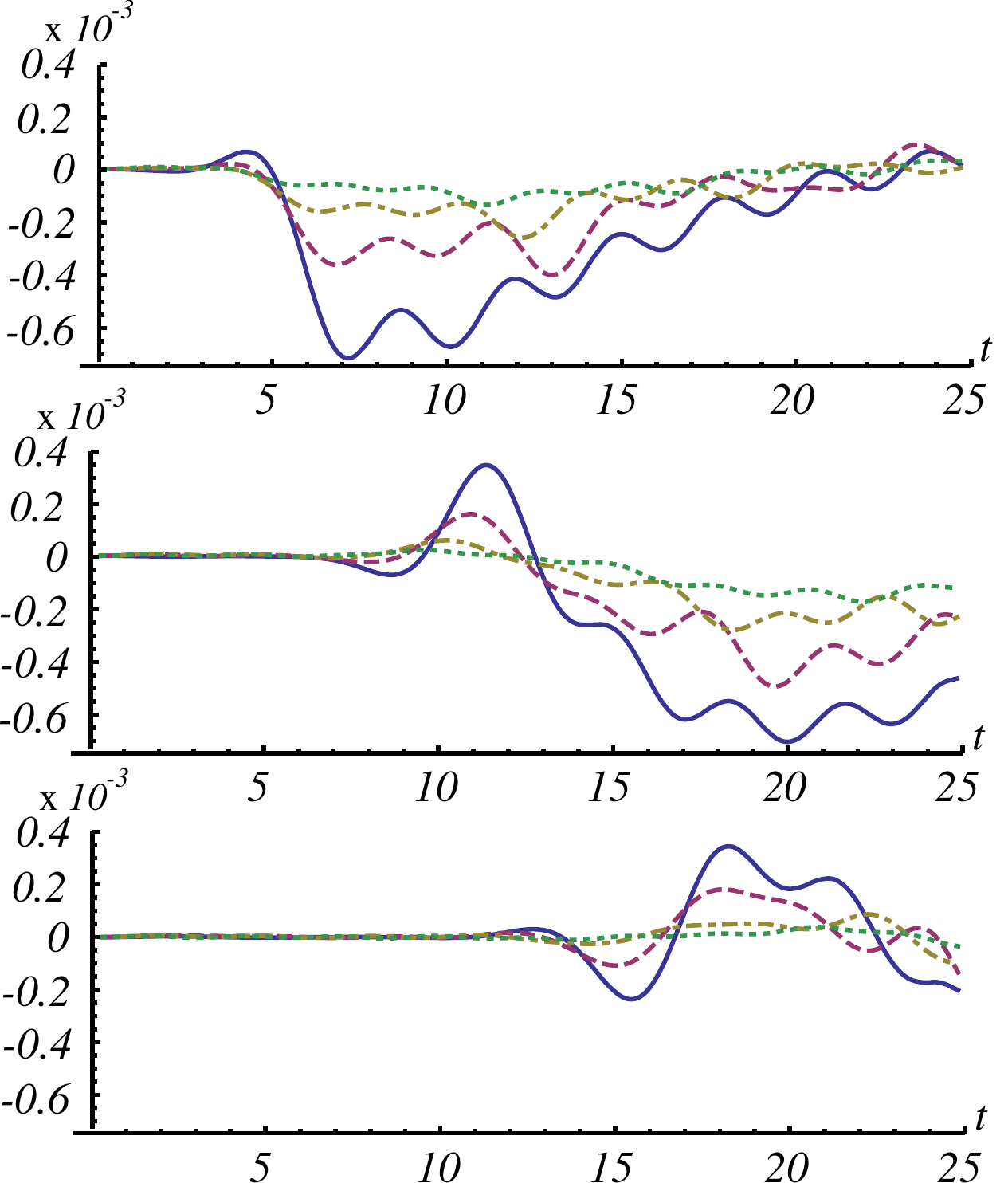}
\caption{(Color online) 
$\Delta{\cal G}_{i}(t,0)$ for $i-i_0=5,10,15,20$, after a 
quench from $U_0=0.025$ to $U=0.03$ as a function of time $t$
for $\lambda=0$ (solid line), $\lambda=0.3$ (dashed line), 
$\lambda=0.6$ (dot-dashed line), {$\lambda=0.9$} (dotted line).}
 \label{fig:detquenchU0.025}
\end{figure}

\begin{figure}[h]
    \includegraphics[width=4.1cm]{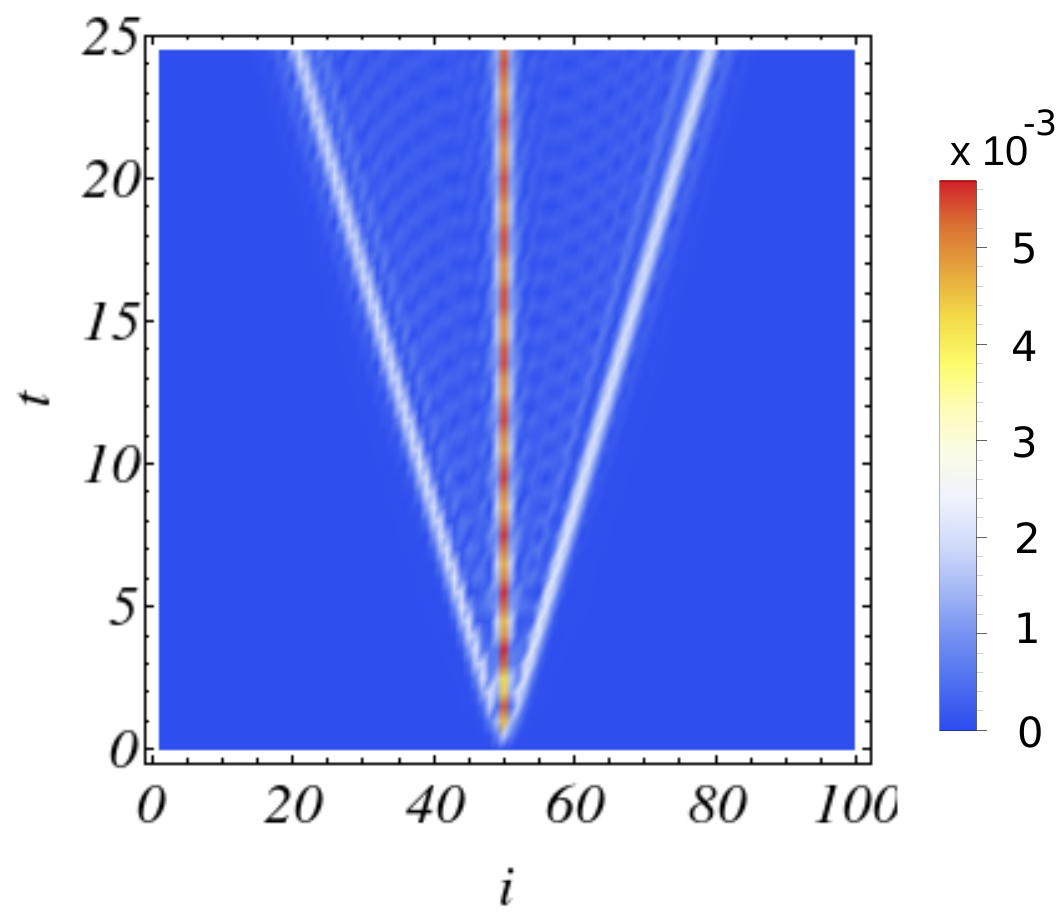}
    \includegraphics[width=4.1cm]{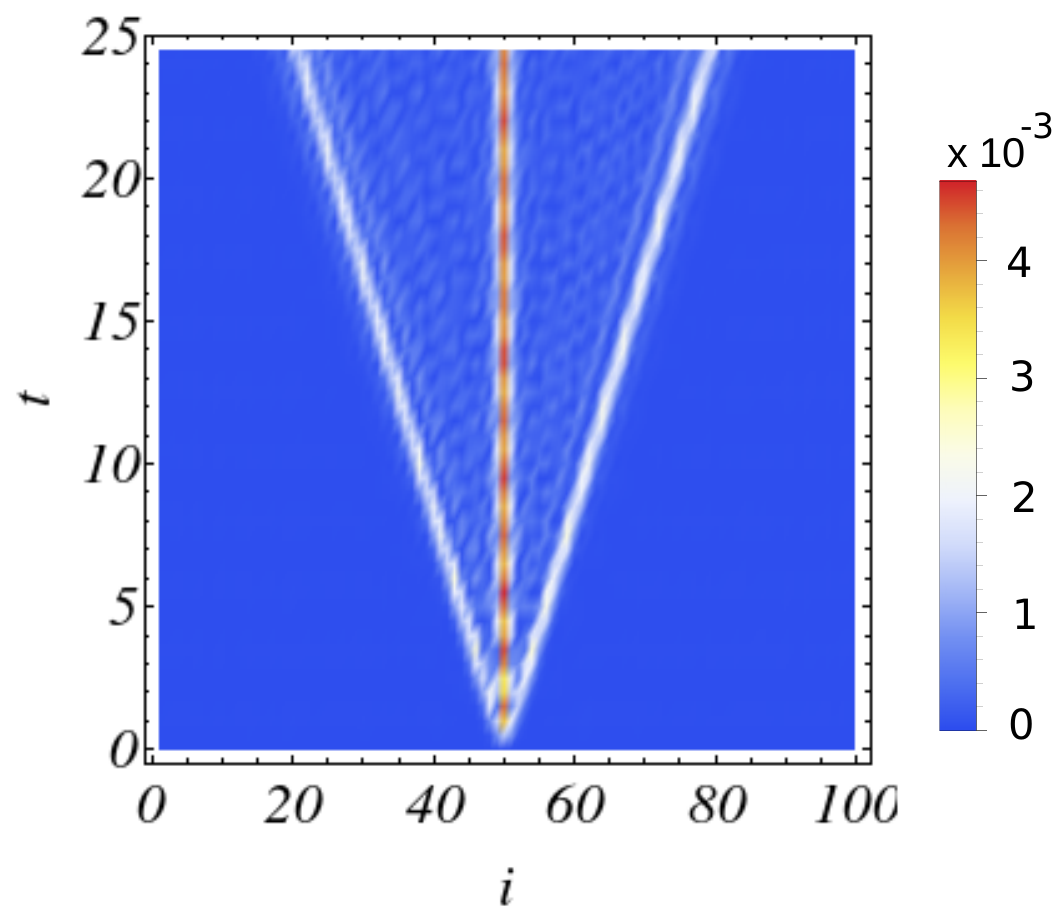}
    \includegraphics[width=4.1cm]{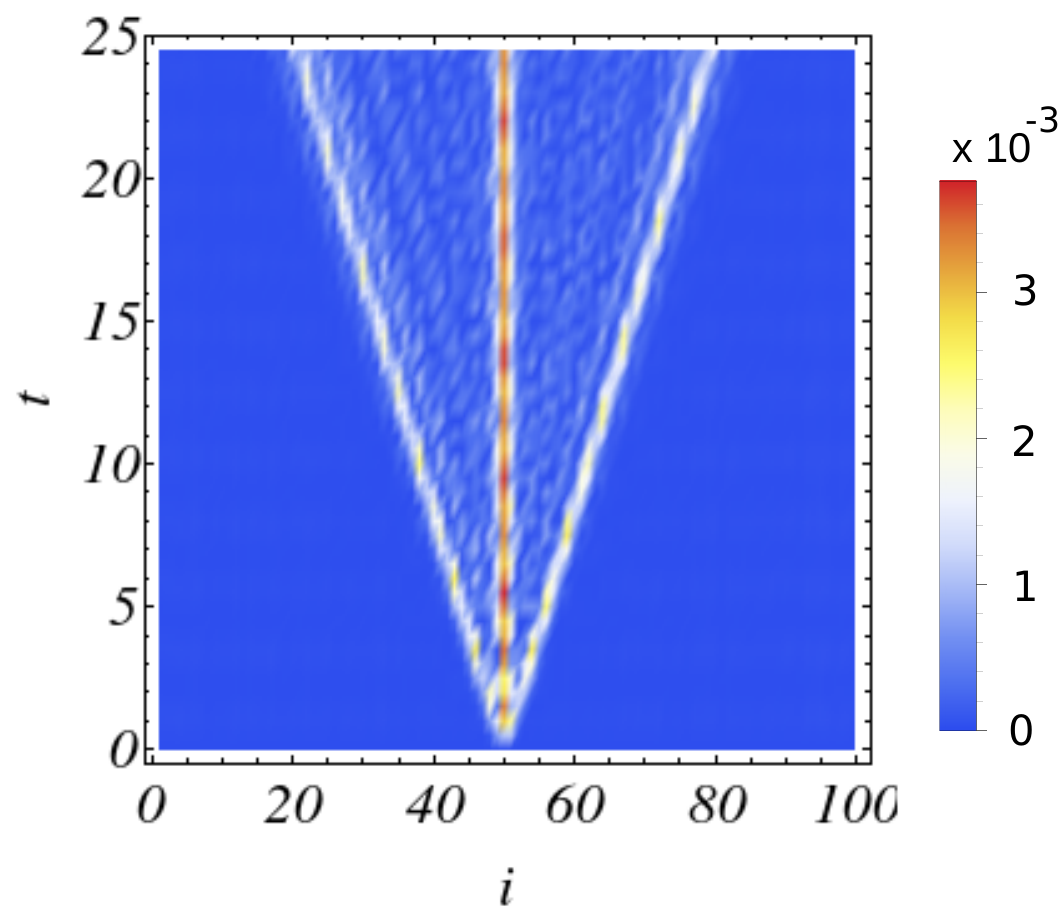}
    \includegraphics[width=4.1cm]{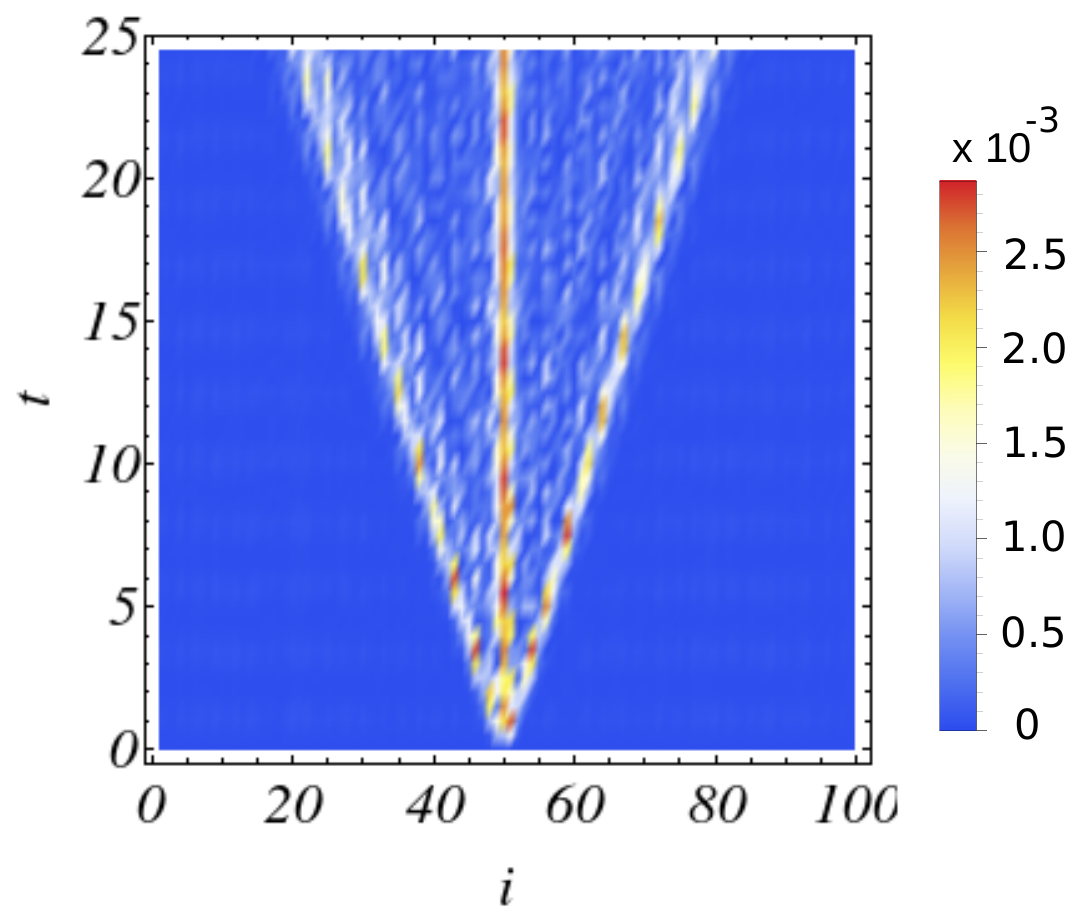}
\caption{(Color online) 
$\Delta{\cal G}_{i}(t,0)$, Eq. (\ref{eq:DGtt}), after a 
quench from $U_0=0.25$ to $U=0.3$, as a function of the 
distance $i$ and time $t$, {for different values of $\lambda$:} 
(from top left to bottom right)
$\lambda=0, 0.3, 0.6, 0.9$.
}
 \label{fig:quenchU0.25}
\end{figure}

\begin{figure}[h]
    \includegraphics[width=6.3cm,height=6.1cm]{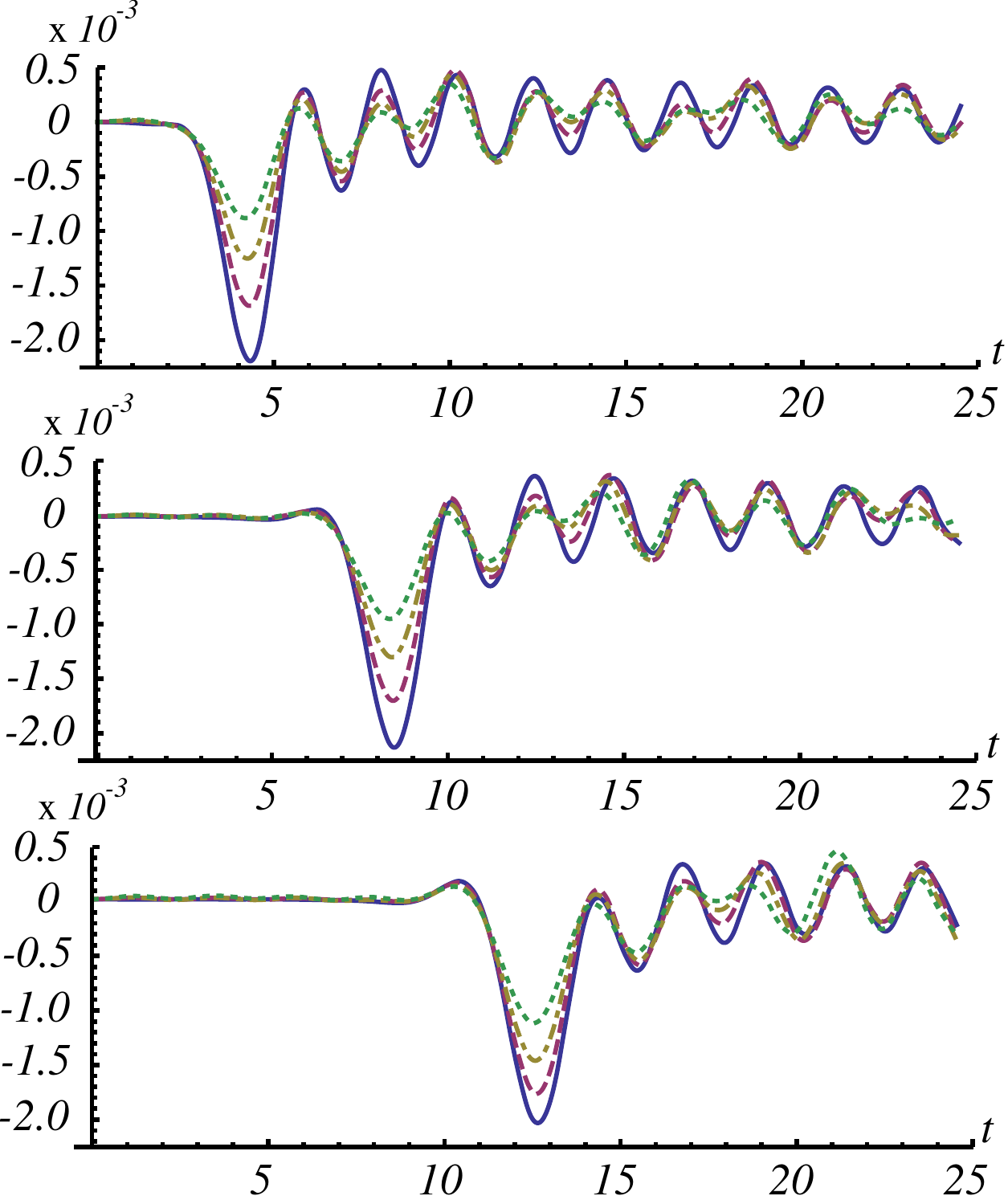}
\caption{(Color online) 
$\Delta{\cal G}_{i}(t,0)$ for $i-i_0=5,10,15,20$, after a 
quench from $U_0=0.25$ to $U=0.3$ as a function of time $t$
for $\lambda=0$ (solid line), $\lambda=0.3$ (dashed line), 
$\lambda=0.6$ (dot-dashed line), $\lambda=0.9$ (dotted line).}
 \label{fig:detquenchU0.25}
\end{figure}

{In Figs. \ref{fig:quenchU0.025} and \ref{fig:detquenchU0.025} 
we plot $\delta {\cal G}_i(t,0)$ after a small quench in a weakly interacting 
system, from $U_0=0.025$ to $U=0.03$, 
for different values of $\lambda$. 
As shown in those plots, the switching on of the Aubry-Andr\'e potential is 
the fate of the fast signals which otherwise would travel ballistically 
at constant velocity given by Eq. (\ref{speed}). 
The spreading is then overall diffusive, although made of rare, sharp and asymmetric timelike signals (see the pattern made of stipes in time, shown in Fig. \ref{fig:quenchU0.025}). 
Increasing $\lambda$ the signals become sparser and sparser, and eventually 
disappear approaching the Bose glass phase. 

In Fig. \ref{fig:quenchU0.25}, we plot $\delta {\cal G}_i(t,0)$ 
for a larger quench in a stronger interacting system, namely} from $U_0=0.25$ to $U=0.3$ for different values of 
$\lambda$. 
{In this case we notice that} the maximum speed at which the 
signals travel, does not depend upon $\lambda$.
This can be seen by looking at the wings of the signal which
have the same width for all values of $\lambda$
and is clearly visible in the plots of Fig. \ref{fig:detquenchU0.25}
by looking at the position of the first peak for different values of
$\lambda$ and different distances.
On the other hand, as $\lambda$ increases the signal 
goes from a purely ballistic dynamics, namely a localized packet
traveling at a constant velocity (see the $\lambda=0$ case)
to a more broadened propagation inside the "light-cone".

\begin{figure}[h!]
 \label{fig:eqtimequenchU0.25}
 \includegraphics[width=4.1cm]{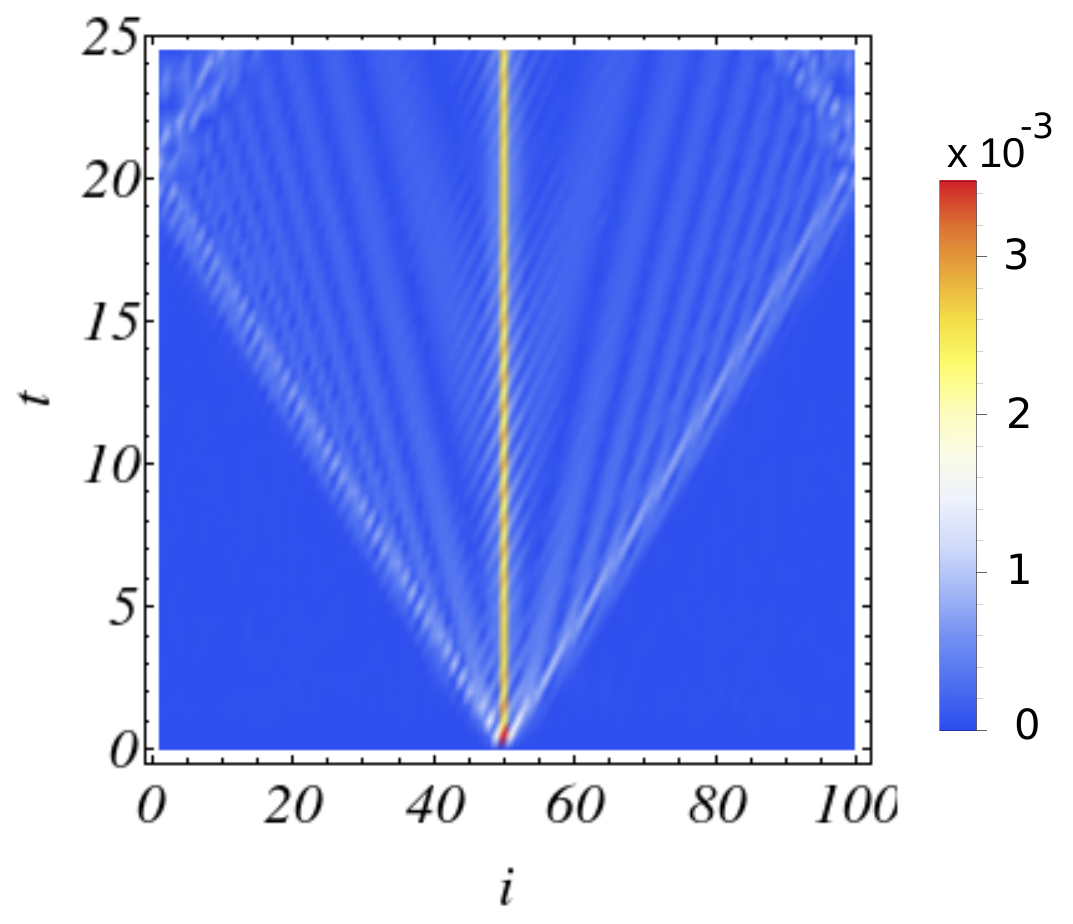}
 \includegraphics[width=4.1cm]{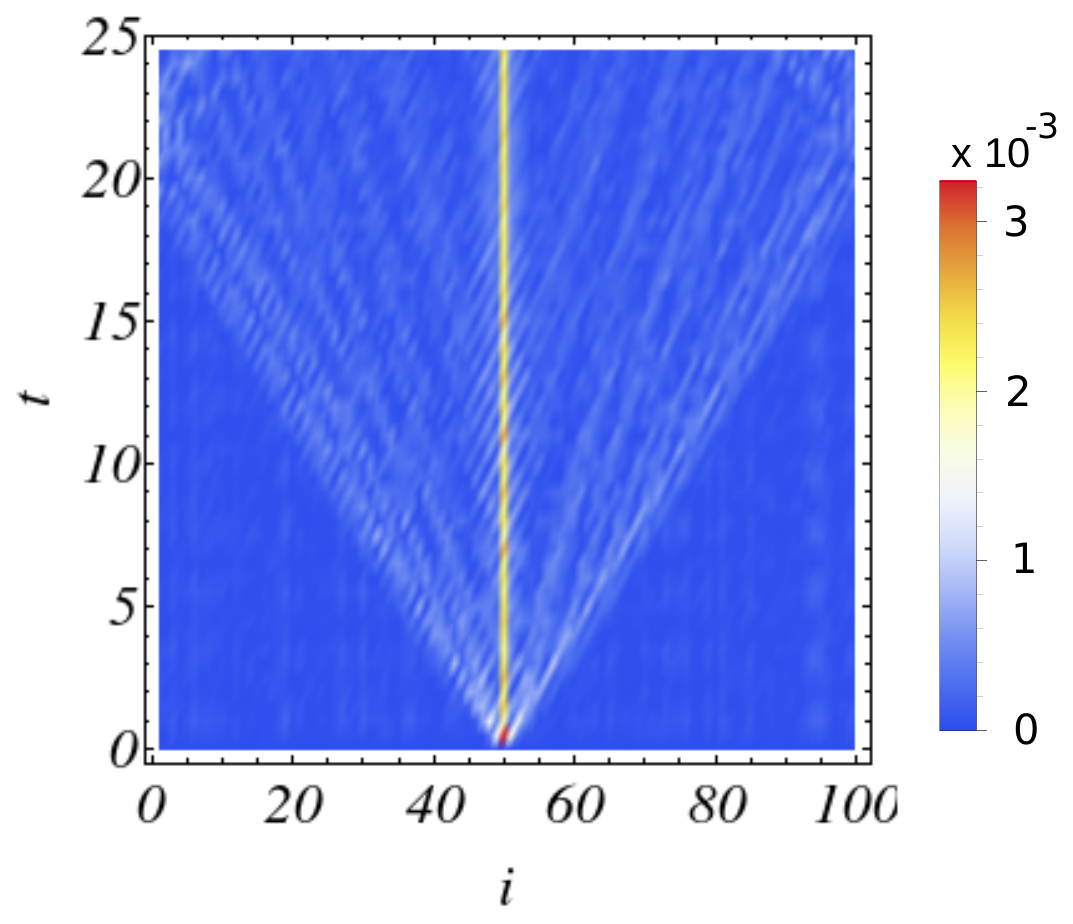}
\caption{(Color online) 
$\Delta{\cal G}_{i}(t,t)$, Eq. (\ref{eq:DGtt}), after a
quench from $U_0=0.25$ to $U=0.3$, as a function of the
distance $i$ and time $t$, for $\lambda=0$ (left), and $\lambda=0.6$ (right).
}
 \label{fig:eqtimequenchU0.25}
\end{figure}

{Finally, let us look at the equal time density-density correlation
function $\Delta {\cal G}_{i}(t,t)$.
As shown in Fig. \ref{fig:eqtimequenchU0.25}, where we plot 
$\Delta{\cal G}_{i}(t,t)$ after a quench in $U$ for two different values of 
$\lambda$, the spreading, whose intensity is much weaker than that of 
$\Delta {\cal G}_{i}(t,0)$, describes a cone with a velocity just 
twice larger than that of the corresponding $\Delta {\cal G}_{i}(t,0)$, 
in particular, in the periodic case ($\lambda=0$), the velocity is $2v$ with 
$v$ given by Eq. (\ref{speed}). Analogously to the different time correlators, 
this velocity is not affected by the aperiodic potential.}

\subsection{Quench in $\lambda$}
\begin{figure}[h]
   \includegraphics[width=4.1cm]{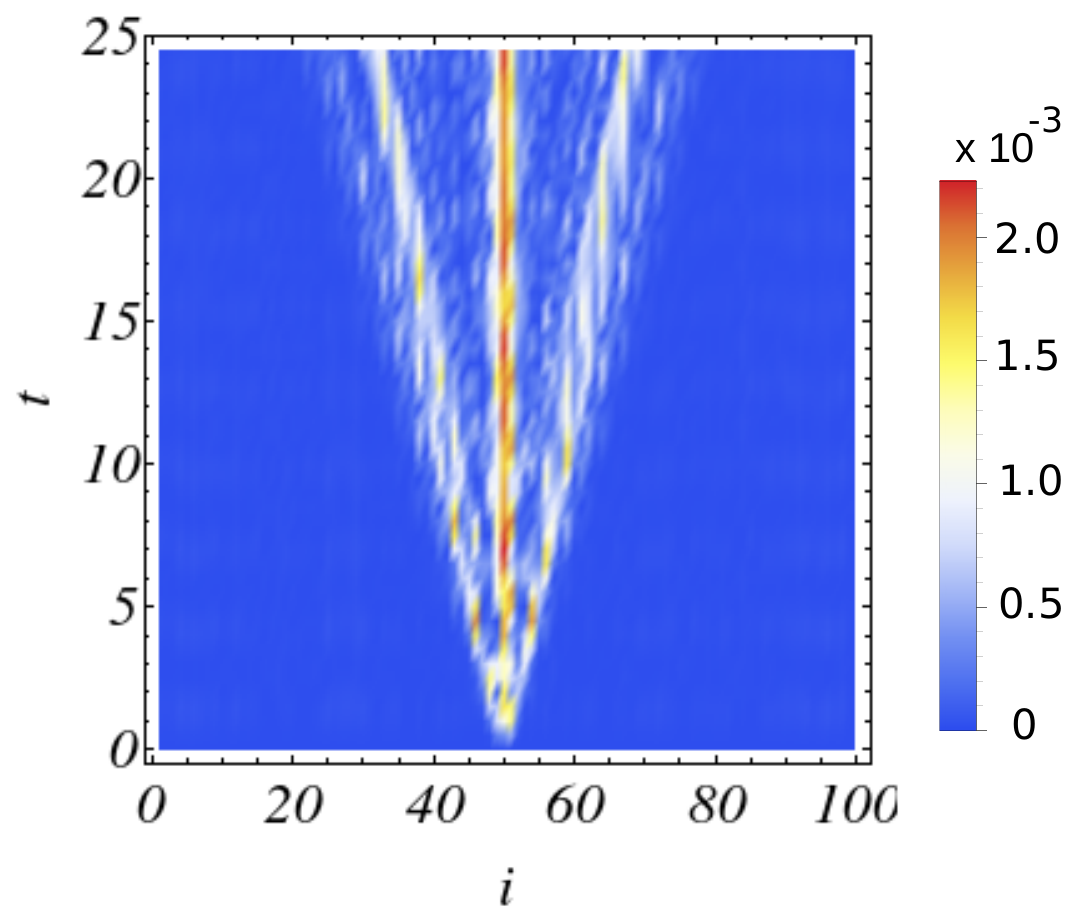}
   \includegraphics[width=4.1cm]{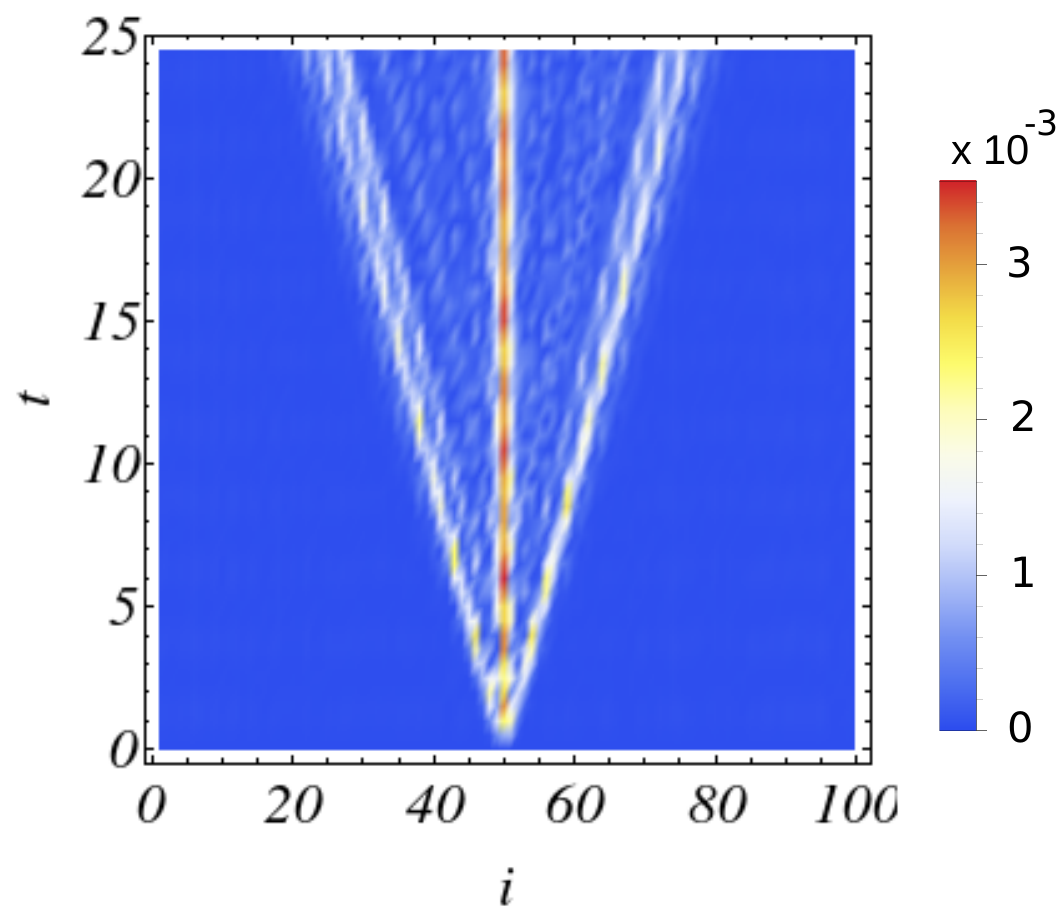}
   \includegraphics[width=4.1cm]{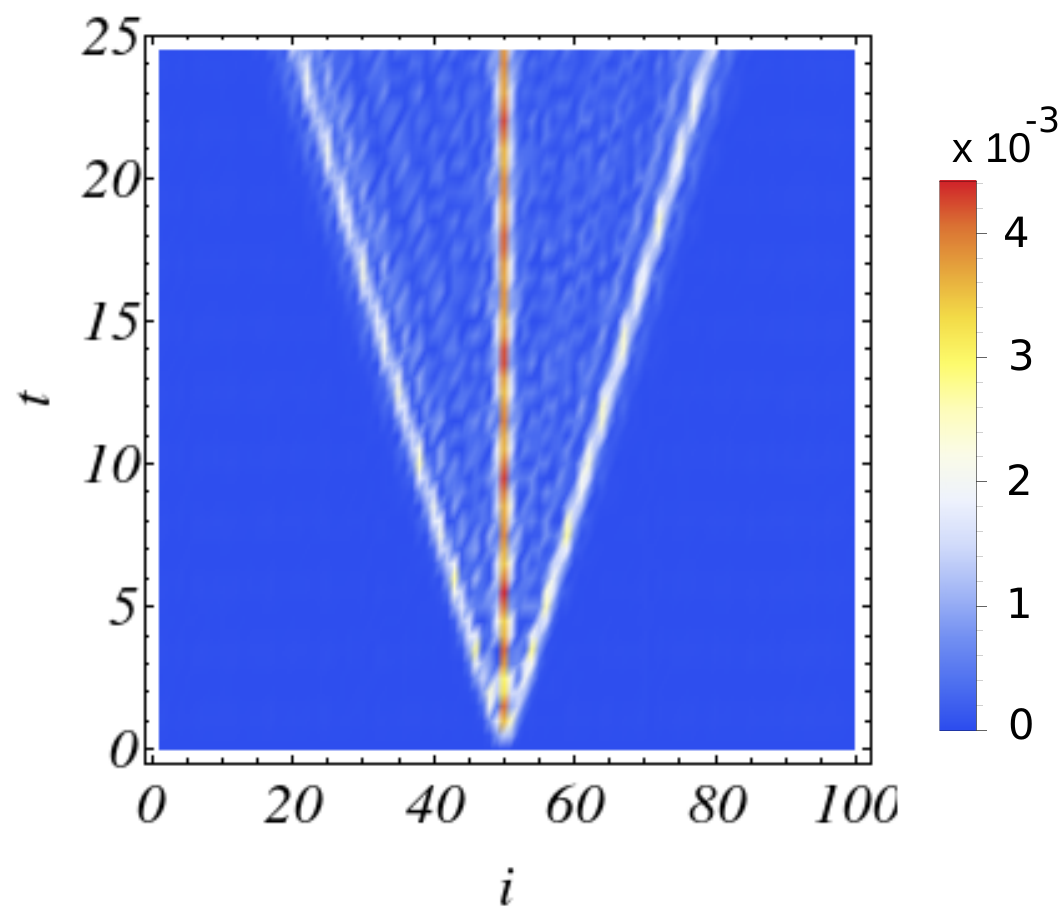}
   \includegraphics[width=4.1cm]{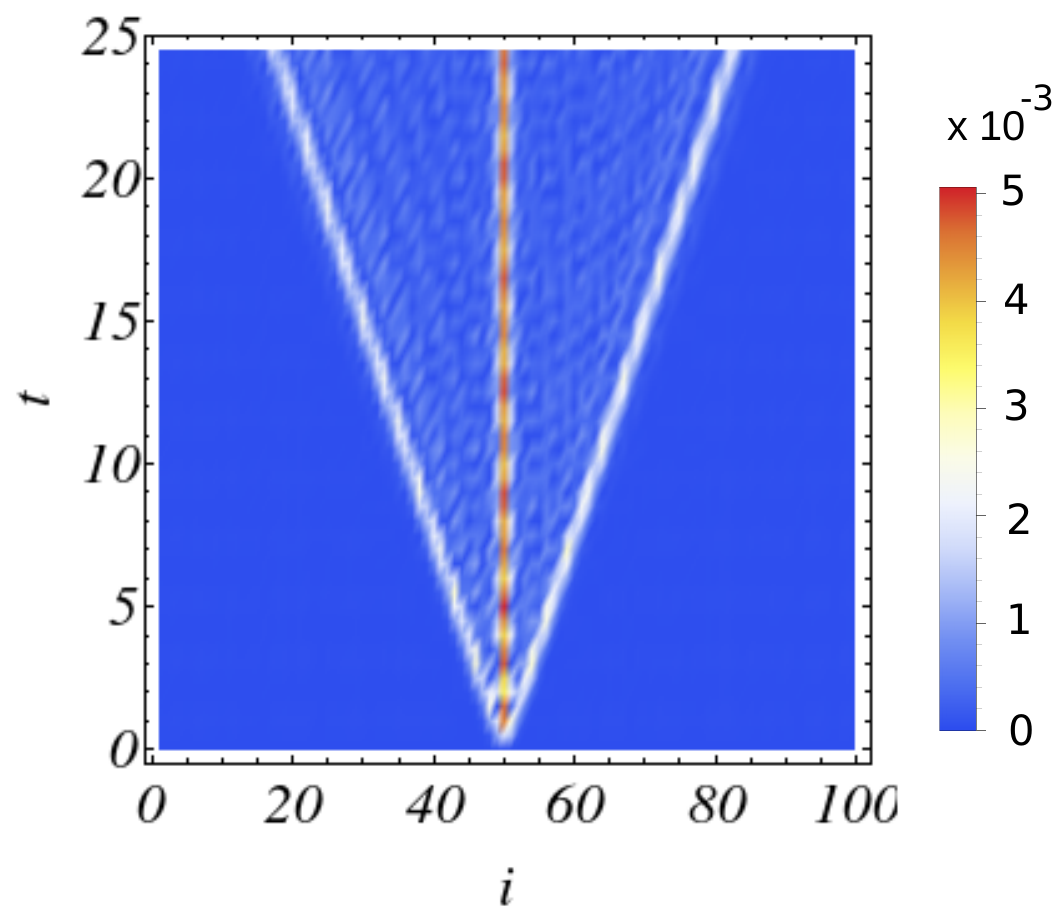}
\caption{(Color online) 
$\Delta {\cal G}_{i}(t,0)$, after a 
quench from $\lambda_0=0.5$ to $\lambda=0.55$, {for different values of $U$}:  
(from top left to bottom right) $U=0.1, 0.2, 0.3, 0.4$.}
 \label{fig:quenchlam0.5}
\end{figure}

\begin{figure}[h]
    \includegraphics[width=6.3cm,height=6.1cm]{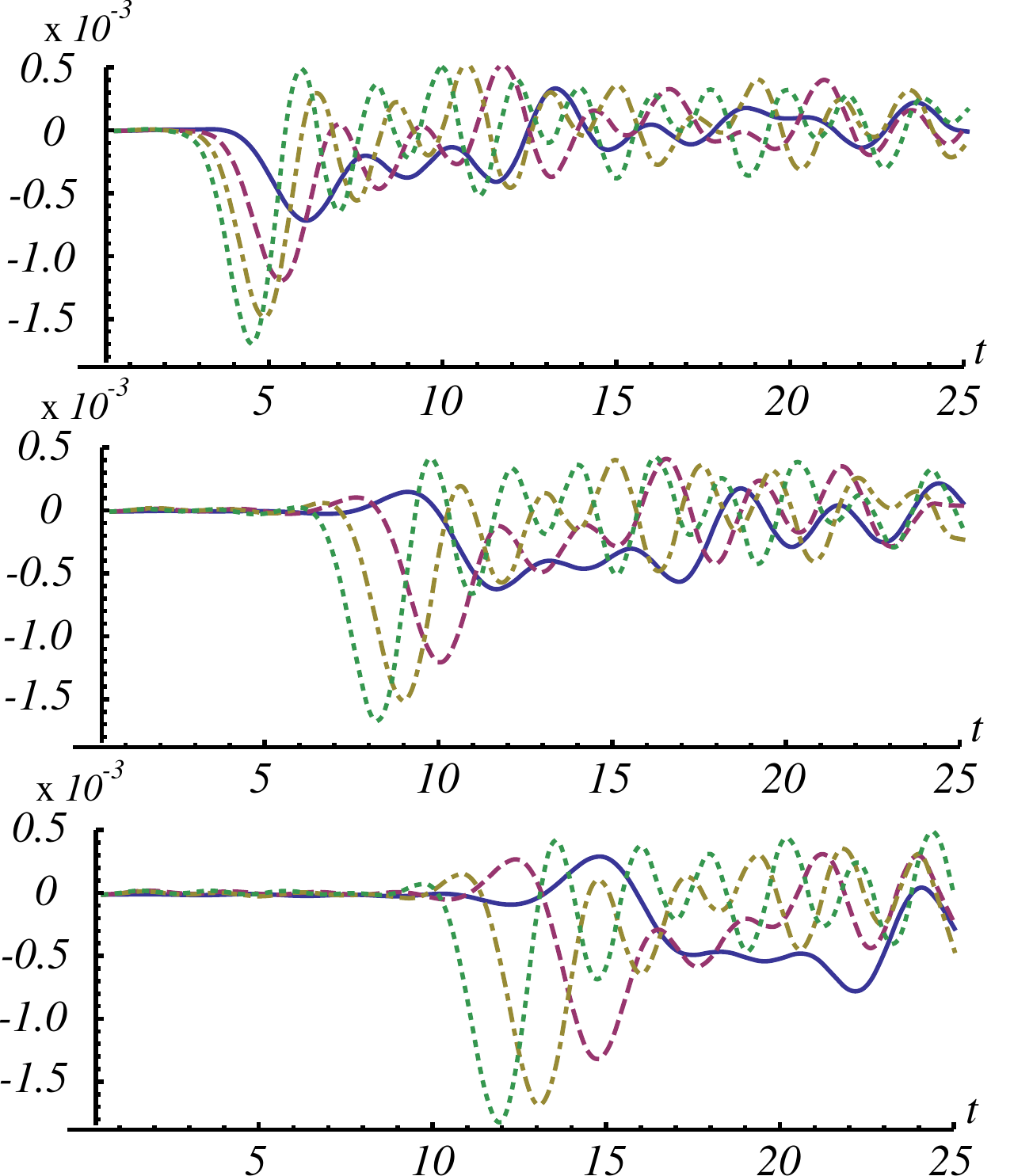}
\caption{(Color online) 
$\Delta{\cal G}_{i}(t,0)$ for $i-i_0=5,10,15,20$, after a 
quench from $\lambda_0=0.5$ to $\lambda=0.55$ as a function of time $t$
for $U=0.1$ (solid line), $U=0.2$ (dashed line), 
$U=0.3$ (dot-dashed line), $U=0.4$ (dotted line).}
 \label{fig:detquenchlam0.5}
\end{figure}

We now consider the case of quenches in the potential
strength $\lambda$ at fixed boson-boson interaction $U$. 
In Fig. \ref{fig:quenchlam0.5} we show the function $\delta {\cal G}_i(t,0)$ 
for a quench from $\lambda_0=0.5$ to $\lambda=0.55$, for different values of the boson-boson interaction,
namely $U=0.1, 0.2, 0.3, 0.4$.
In this case we can clearly see an increase in the speed propagation
of the signal as $U$ increases (widening of the outermost wings), 
better visible in Fig. \ref{fig:detquenchlam0.5} where 
the signal appears at {early} times as $U$ is increased
for a given distance from $i_0$. 
This is in agreement with the fact that, 
{in the homogeneous system, an increase in the boson-boson interaction 
would lead to an increase of the group velocity. For large $U$, therefore, the Aubry-Andr\'e potential becomes marginal even if the dynamics is generated 
by a sudden quench of $\lambda$.  
Moreover, increasing $U$ we notice that the slower diffusive signals become weaker and weaker exhibiting a crossover to an almost pure ballistic expansion.}

\section{Momentum distribution}
{In this section we look at the (time dependent) momentum distribution 
defined as the Fourier transform of the one-body density matrix}
\begin{eqnarray}
 \label{eq:structfact}
 n(k,t)&=&\frac{1}{L}\sum\limits_{i,j}e^{-\imath k (i-j)} 
\left\langle \hat{b}^\dagger_i(t)\hat{b}_j(t)\right\rangle\\
\nonumber&=&n_0(k)+n_{ex}(k,t)
\end{eqnarray}
{with $\hat b$ given by Eq. (\ref{b}),   
$n_0(k)=\frac{N_0^2}{L} \sum_{i,j}e^{-\imath k (i-j)} \phi_i^*\phi_j$, 
is the mean field contribution, and 
$n_{ex}(k,t)=\frac{1}{L}\sum\limits_{i,j}e^{-\imath k (i-j)}
\langle \hat{c}^\dagger_i(t)\hat{c}_j(t)\rangle$
the fluctuation contribution.}

\begin{figure}[h]
   \includegraphics[width=7cm]{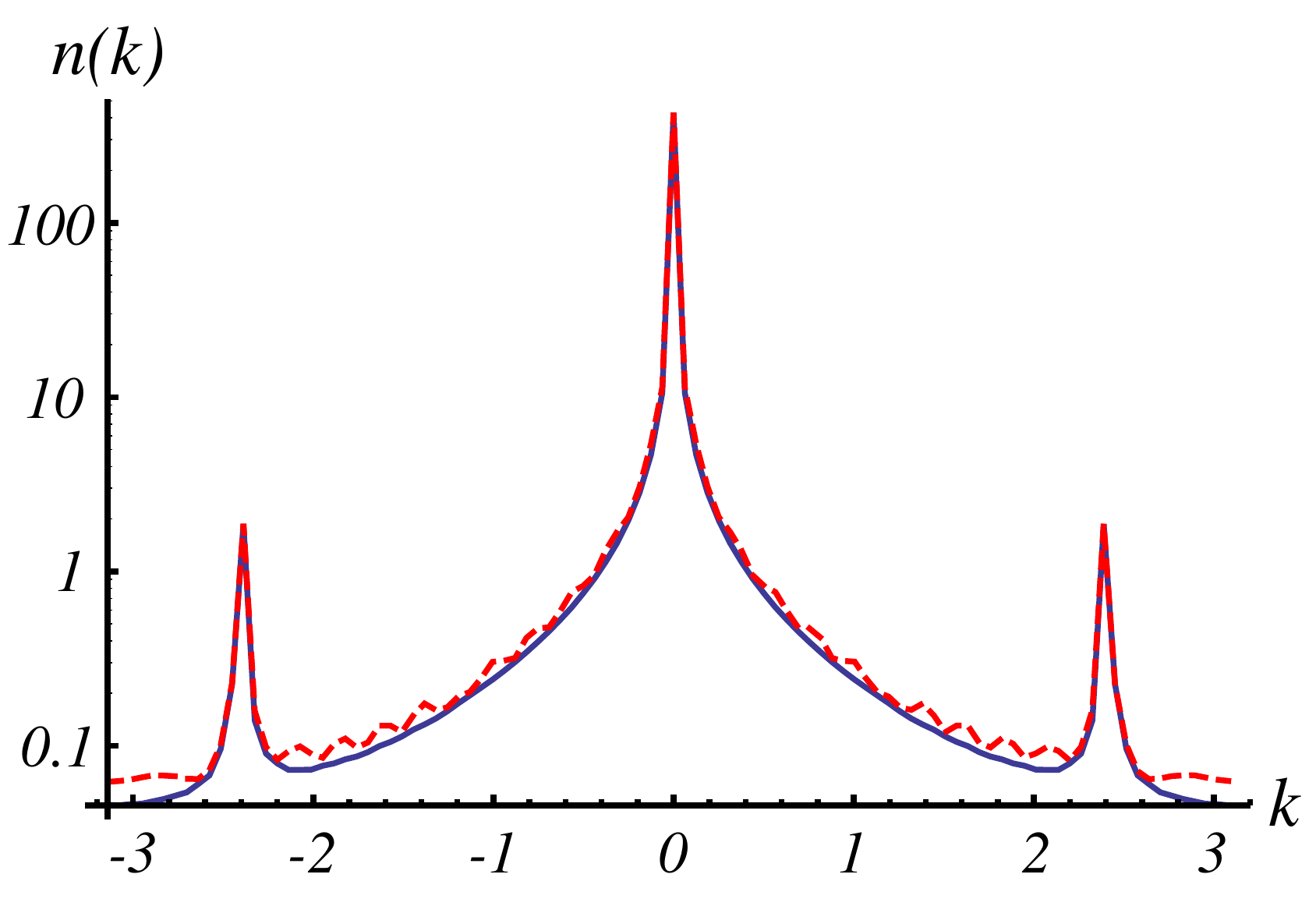}
\caption{(Color online) 
Momentum distribution $n(k,t)$, Eq. (\ref{eq:structfact}) with $L=100$, {at $t=0$ (solid blue line) and at $t=12.5$ (dashed red line) after a quench from $U_0=0.25$ to $U=0.3$.}}
 \label{fig:S}
\end{figure}

{In Fig. \ref{fig:S} we plot the momentum distribution $n(k,t)$ 
at two different 
times, at $t=0$ and at later time after a sudden change of the boson-boson interaction $U$.
Three peaks are clearly visible at $k=0, \pm 2\pi(1-\tau^{-1})$ 
due to the presence of the modulation of the potential,  
in agreement with DMRG (density matrix renormalization group) 
calculations reported  
in Refs. \cite{roux2008,deng2008}, where it 
was shown that 
peaks appear at $k=\pm 2\pi(1-r)$, if the on-site potential has the functional
form $\cos(2\pi r i)$. In our case $r=\tau=1+\tau^{-1}$ and thus $\cos(2\pi 
\tau i)=\cos(2\pi i/\tau)$.}
{The quantum quench in $U$ weakly modifies the profile of the momentum distribution, at least in the time scale considered, inducing a small 
modulation due to quantum fluctuations. This result suggests that one has to rather focus on the density-density correlations for better detecting the effects of quench dynamics.}

\begin{figure}[h!]
\begin{tabular}{cc}
\hspace{-0.4cm}
\includegraphics[width=4.3cm]{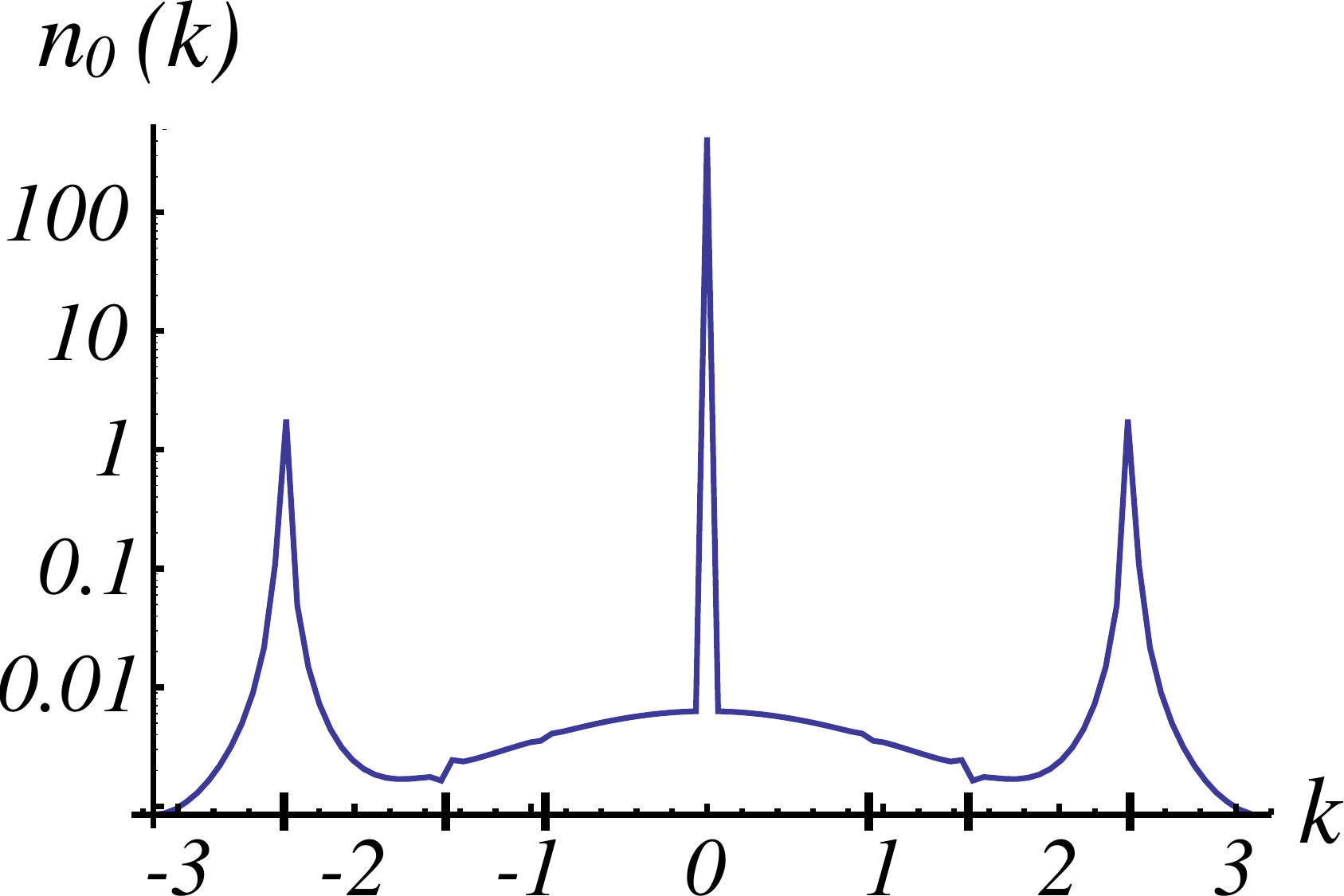}
 \includegraphics[width=4.3cm]{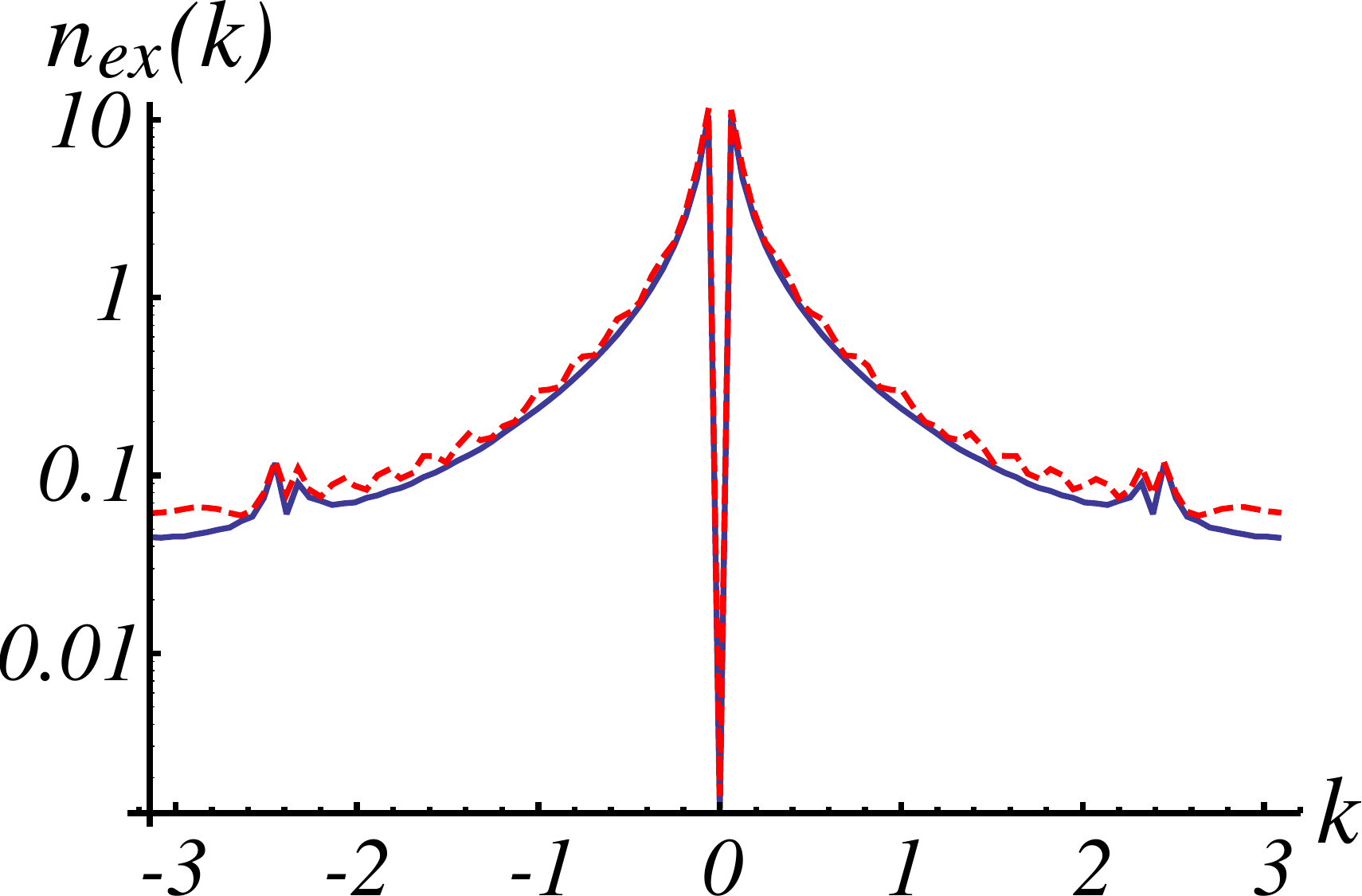}
\end{tabular}
\caption{(Color online) 
Mean field ($n_0$) and fluctuation ($n_{ex}$) contributions to the momentum distribution {for $\lambda=0.6$, at $t=0$ (solid blue line) and at $t=12.5$ 
(dashed red line) after a sudden quench from $U_0=0.25$ to $U=0.3$. }
Large ticks on the $k$-axis of the plot for $n_0(k)$, highlight features at $k=\pm2\pi(1-\tau^{-1})/\tau^\ell$ with $\ell=0,1,2$.}
 \label{fig:S0}
\end{figure}

{Looking carefully at the mean field term, $n_0(k)$ 
(see Fig. \ref{fig:S0}), 
 we notice other peculiar features due to the scaling propeties of the Aubry-Andr\'e potential, 
at positions $k=\pm2\pi(1-\tau^{-1})/\tau^\ell$ with $\ell=0,1,2$, 
while the contribution due to fluctuations $n_{ex}(k,t)$ (Fig. \ref{fig:S0}, plot on the right) exhibits dips at $k=0, \pm 2\pi(1-\tau^{-1})$ at any time.
This ensures that the time dependent Bogoliubov approach is consistent with 
the assumption that the mean field dynamics can be neglected on the time scales considered.}

\section{Loschmidt echo}
\label{sec:lodecho}
In this section we calculate the vacuum persistence amplitude following a 
sudden quench, defined as
\begin{equation}
 \label{nu1}
 \nu(t)=\left\langle e^{i \hat H_0 t}\, e^{-i\hat H t}\right\rangle ,
\end{equation}
where the average is over the initial state, $|\psi(0)\rangle$, namely 
the vacuum state for $\hat\alpha_n$, $\hat\alpha_n|\psi(0)\rangle=0$. 
It is, therefore, convenient to define

\be
\delta{\hat H}=\hat H-\hat H_0
\ee
so that we can rewrite Eq. (\ref{nu1})
\bea
\label{nu2}
&& \nu(t)=\left\langle {\mathcal T}e^{-i\int_{0}^{t}d\tau \,\delta \hat H(\tau)}\right\rangle\\
&&\nonumber\phantom{\nu(t} 
=\sum_{n=0}^\infty \frac{(-i)^n}{n!}\int_0^t d\tau_1...\int_0^t d\tau_n\,\langle{\mathcal T}[\delta\hat H(\tau_1)...\delta \hat H(\tau_n)]\rangle
\eea
where $\delta \hat H(\tau)=e^{i\hat H_0 \tau}\,\delta \hat H \,e^{-i\hat H_0 \tau}$, 
in the interaction picture with respect to the Hamiltonian $\hat H_0=\sum_n \epsilon^0_n\hat\alpha^\dag_n\hat\alpha_n$. \\
For simplicity, calling $h_{ij}=(V_i-\mu+2g|\phi_i|^2)\delta_{ij}-J\delta_{j,i\pm 1}$ and $\Delta_i=g \phi_i^2/2$, we rewrite Eq. (\ref{eq:Hb}) as 
\be
\hat H=\sum_{i,j} h_{ij}\,\hat c^\dag_i \hat c_j +\sum_i(\Delta_i \,\hat c^\dag_i 
\hat c^\dag_i+\Delta^*_i\, \hat c_i \hat c_i).  
\ee
Analogously, we can rewrite $H_0$ with $h^0_{ij}$ and $\Delta^0_i$, and $\delta \hat H$ 
with $\delta h_{ij}=h_{ij}-h^0_{ij}$ and $\delta\Delta_i=\Delta_i-\Delta^0_i$. By applying the Bogoliubov transformation in Eq.(\ref{bog1}), 
one can write $\delta \hat H$ in terms of the initial Bogoliubov operators, $\alpha_n$, which in the interaction picture can be written as
\bea
\label{dH}
&&\delta \hat H(\tau)=E_0+\sum_{n,\ell} A_{n\ell}\,e^{i(\epsilon^0_n-\epsilon^0_\ell)\tau}
\hat\alpha^\dag_n\hat\alpha_\ell\\
\nonumber&&\phantom{\delta H}
+\sum_{n,\ell}\left(B_{n\ell}\,e^{-i(\epsilon^0_n+\epsilon^0_\ell)\tau}
\hat\alpha_n\hat\alpha_\ell+B^*_{n\ell}\,e^{i(\epsilon^0_n+\epsilon^0_\ell)\tau}
\hat\alpha^\dag_\ell\hat\alpha^\dag_n\right)
\eea
where the constant term
\[
E_0=\sum_\ell\Big(\sum_{i,j} \delta h_{ij}\,v_{i,\ell}v^*_{j,\ell}-2\sum_i\textrm{Re}\left[\delta\Delta_i\, u^*_{i,\ell}v_{i,\ell}\right]\Big)
\]
is just a phase shift in Eq. (\ref{nu2}), while 
\bea
\nonumber
&&A_{n\ell}
=\sum_{i,j}\delta h_{ij}(u^*_{i,n}u_{j,\ell}+v_{i,\ell}v^*_{j,n})\\
&&\phantom{A_{n\ell}}
-2\sum_i\textrm{Re}\left[\delta\Delta_i(u^*_{i,n}v_{i,\ell}+v_{i,\ell}u^*_{i,n})\right]
\\
\nonumber
&&B_{n\ell}
=\sum_i(\delta\Delta_i v_{i,n}v_{i,\ell}+\delta\Delta^*_i u_{i,n}u_{i,\ell})
-\sum_{i,j}\delta h_{ij} v_{i,n} u_{j,\ell}\\
\eea
Notice that, since $\delta h_{ij}=\delta h_{ji}$, then $A_{n\ell}=A^*_{\ell_n}$, as should be in order for $\delta \hat H$ to be hermitian. Moreover, in Eq. (\ref{dH}), because of commutation relations, only the symmetric part of $B_{n\ell}$, namely $(B_{n\ell}+B_{\ell n})/2$, plays a role, analogously for $B^*_{n\ell}$. \\
Now, exploiting the linked cluster expansion theorem, we get 
\begin{equation}
 \label{lce}
 \ln{\nu(t)}={-iE_0t}+{\sum_{q=1}^\infty{\cal C}_q(t)}
\end{equation}
where ${\cal C}_q=(-i)^q\int_0^t d\tau_1...\int_0^{\tau_{q-1}} d\tau_q\,\langle
\delta^\prime \hat H(\tau_1)...\delta^\prime \hat H(\tau_q)\rangle_{c}$ is the 
sum of all connected diagrams of the $q$-th 
order in the perturbation parameters $\delta h_{ij}$, $\delta\Delta_i$ and where $\delta' \hat H=\delta \hat H-E_0$.
In what follows we will consider diagrams up to third order.
After time integration, we get (${\cal C}_1(t)=0$) 
\begin{widetext}
\begin{eqnarray}
 \label{C2}
 {\cal C}_2(t)&=&2i\sum_{n,\ell}\frac{|B_{n \ell}|^2}
{(\epsilon^0_{n}+\epsilon^0_{\ell})} \;t
-2\sum_{n,\ell}\frac{|B_{n \ell}|^2}
{(\epsilon^0_{n}+\epsilon^0_{\ell})^2}
\left(1-e^{-i(\epsilon^0_n+\epsilon^0_{\ell})t}\right)\\
 \label{C3}
 {\cal C}_3(t)&=&-4i\sum_{n,\ell,m}\frac{B^*_{n \ell}\,B_{\ell m}\,A_{m n}}{(\epsilon^0_{n}+\epsilon^0_{\ell})(\epsilon^0_{\ell}+\epsilon^0_{m})}\;t
+4\sum_{n,\ell, m}\frac{B^*_{n \ell}\,B_{\ell m}\,A_{m n}}{\epsilon^0_{n}-\epsilon^0_{m}}\left[
\frac{1-e^{-i(\epsilon^0_{\ell}+\epsilon^0_{m})t}}{(\epsilon^0_{\ell}+\epsilon^0_{m})^2}
-\frac{1-e^{-i(\epsilon^0_{n}+\epsilon^0_{\ell})t}}{(\epsilon^0_{n}+\epsilon^0_{\ell})^2}\right]
\end{eqnarray}
\end{widetext}
One can easly show, by the properties os the coefficients $A_{n\ell}$ and $B_{n,\ell}$, 
that the first terms of Eqs. (\ref{C2}), (\ref{C3}) are purely imaginary.
However, without loss of generality, since $u_{i,n}$, $v_{i,n}$ and $\phi_i$ 
can be chosen to be real, then also $A_{n\ell}$ and $B_{n\ell}$ can be real.

In the following we will look at the behavior of the Loschmidt echo defined 
as 
\be
{\cal L}(t)=|\nu(t)|^2=e^{\sum_q 2\textrm{Re} [{\cal C}_q(t)]}
\ee
{after a quantum quench in the interaction ($U$) or in the potential ($\lambda$) parameters.} 

\subsection{Periodic case, $\lambda=0$}
As a reference, let us first consider the homogeneous case ($\lambda=0$). 
In this case only an interacting quench can be made ($\delta U\neq 0$), and, 
keeping for simplicity only the first non-vanishing contribution, 
Eq. (\ref{C2}), dominant for small $\delta U$, we get 

\be
\label{C2_hom}
\textrm{Re}[{\cal C}_2]=-\frac{\delta g^2}{4}\sum_k 
\left(\frac{\sin(\sqrt{\varepsilon_k(\varepsilon_k+2U_0\nu_0)}\,t)}
{\varepsilon_k+2\nu_0 U_0}\right)^2
\ee
where $\varepsilon_k=J(1-\cos k)$ is the single particle dispersion, $\nu_0=N_{00}/L$ the condensate density at $t=0$, $U_0$ the initial value of the interaction parameter, and finally $\delta g=g-g_0=UN_0-U_0N_{00}$. A rought evaluation of Eq. (\ref{C2_hom}) can be obtained expanding $\varepsilon_k\simeq Jk^2/2$, so that 
$\textrm{Re}[{\cal C}_2]\simeq -\frac{\delta g^2}{8\pi}\frac{\pi(1-e^{-4 U_0\nu_0 t}(1+4U_0\nu_0 t)}{16\sqrt{J}(U_0\nu_0)^{3/2}}$, finding that, for large time, the Loschmidt echo saturates at the value
\be
{\cal L}\left(t\gtrsim (U_0\nu_0)^{-1}\right)\propto  \exp\left[-\frac{\delta g^2}{(U_0 \nu_0)^{{3}/{2}}}\right]
\label{Lsaturation}
\ee
At later time further corrections may play a role
and the third order diagrams need to be included.

\subsection{Quenches in $U$}
\begin{figure}[h]
\begin{center}
  \includegraphics[width=5.9cm]{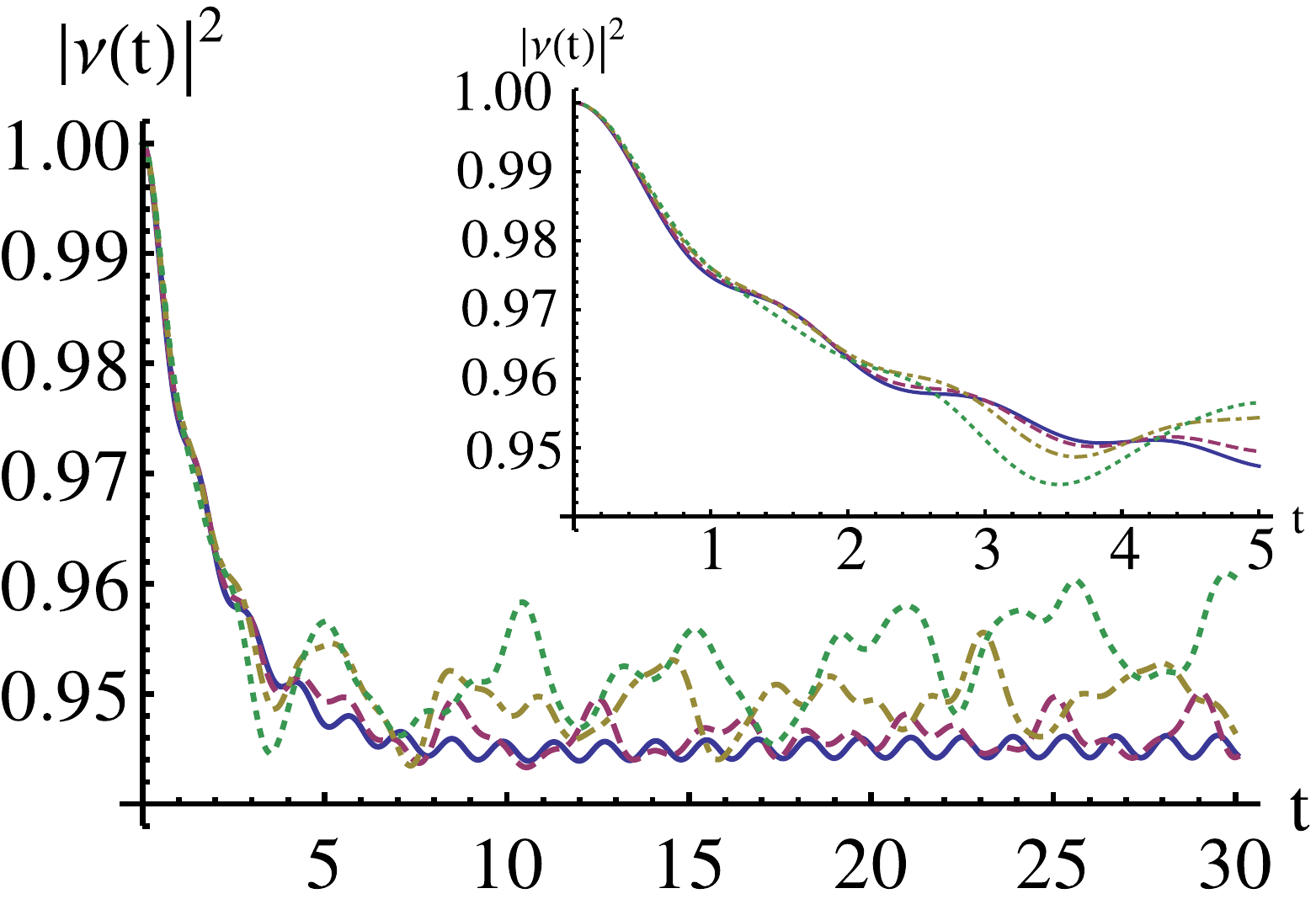}
  \includegraphics[width=5.9cm]{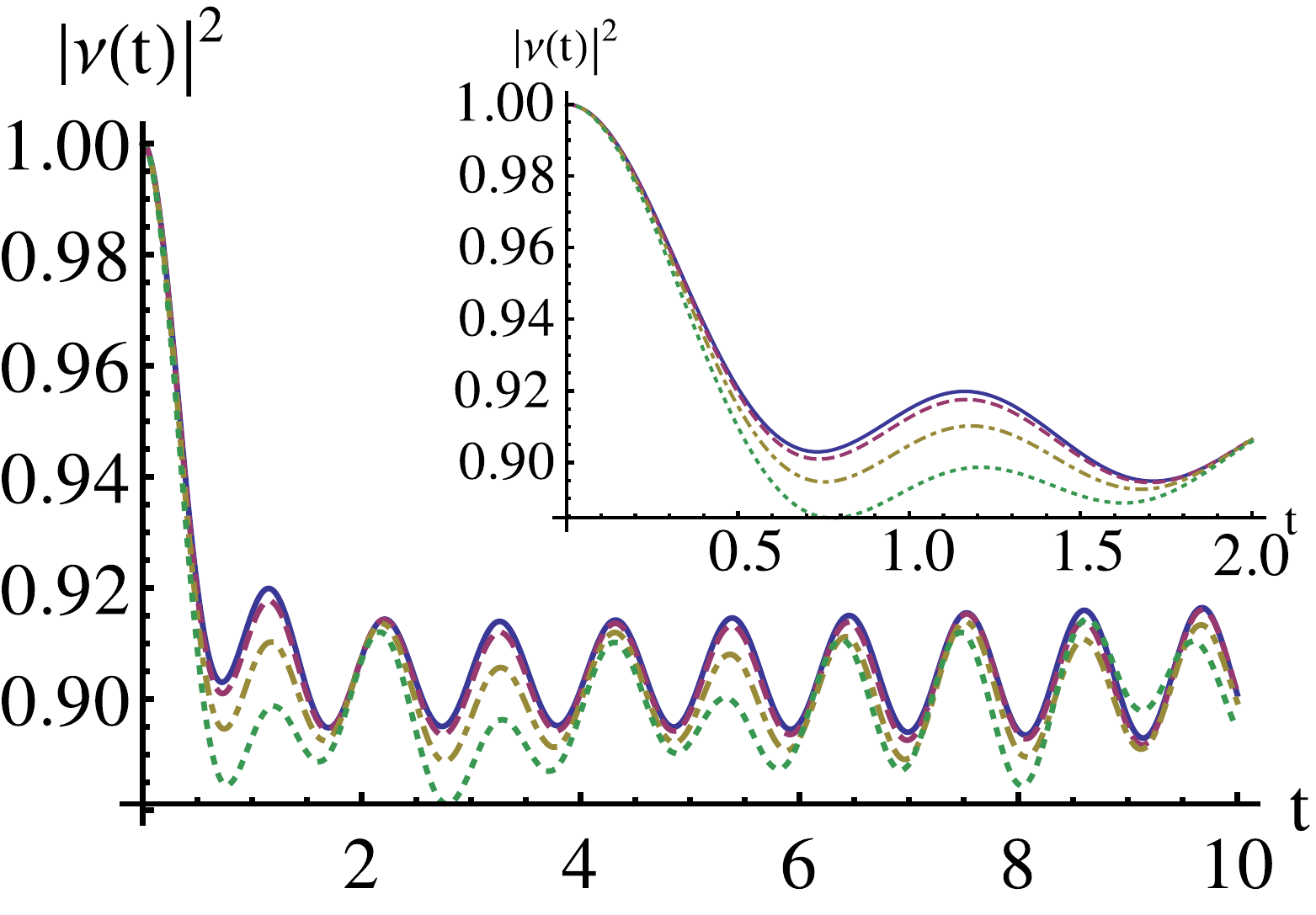}
\caption{(Color online) 
${\cal L}(t)$, after a 
quench (top) from $U_0=0.05$ to $U=0.06$ and  
(bottom) from $U_0=0.25$ to $U=0.3$, {for different values of 
$\lambda=\lambda_0$}: 
$\lambda=0$ (solid line), $\lambda=0.3$ (dashed line),
$\lambda=0.6$ (dot-dashed line), $\lambda=0.9$ (dotted line).
In the insets the short time dynamics of the echo showing the 
characteristic quadratic decay at short times.}
 \label{fig:echoquenchU}
\end{center}
\end{figure}

In the case of a quench in $U$ the echo shows 
a quadratic decay at short times 
and an exponential decay on longer time scales
approaching {a stationary value approximatelly given by Eq. 
(\ref{Lsaturation})}, which however does not correspond to the 
overlap between the initial vacuum state and the one of the final Hamiltonian. 
Oscillations around this {stationary} value are induced by the 
bandwidth. {At finite $\lambda$, the presence of additional sub-bands, 
as shown in Fig. \ref{fig:spectr}, induces further low frequencies in 
the echo. 
From Fig. \ref{fig:echoquenchU}}, we see that also for moderately 
large quench amplitudes
the echo is always close to one, meaning that 
a quench in $U$ does not make the system to fully explore the phase space
and the system stays close to its initial state.\\
The effect of $\lambda$ is to make the system more chaotic 
further reducing the overlap between the initial and the time evolved
state for large $U$. 
{Remarkably, at low $U$, the Loschmidt echo, at $t>(U\nu)^{-1}$, increases as $\lambda$ is increased, as shown in the first plot of Fig. \ref{fig:echoquenchU}.}

%
%
\begin{figure}[h!]
 \includegraphics[width=5.86cm]{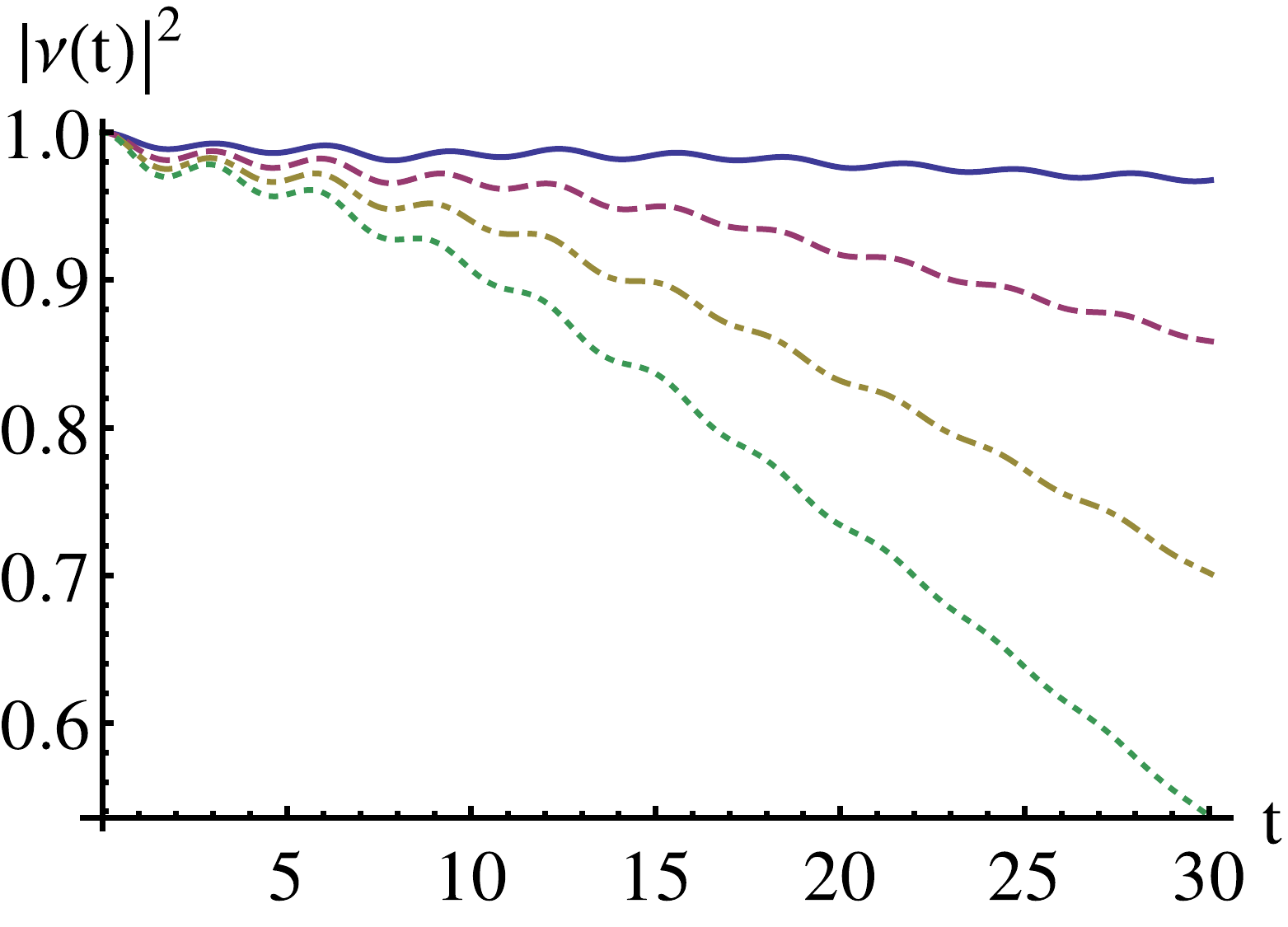}
  \includegraphics[width=5.86cm]{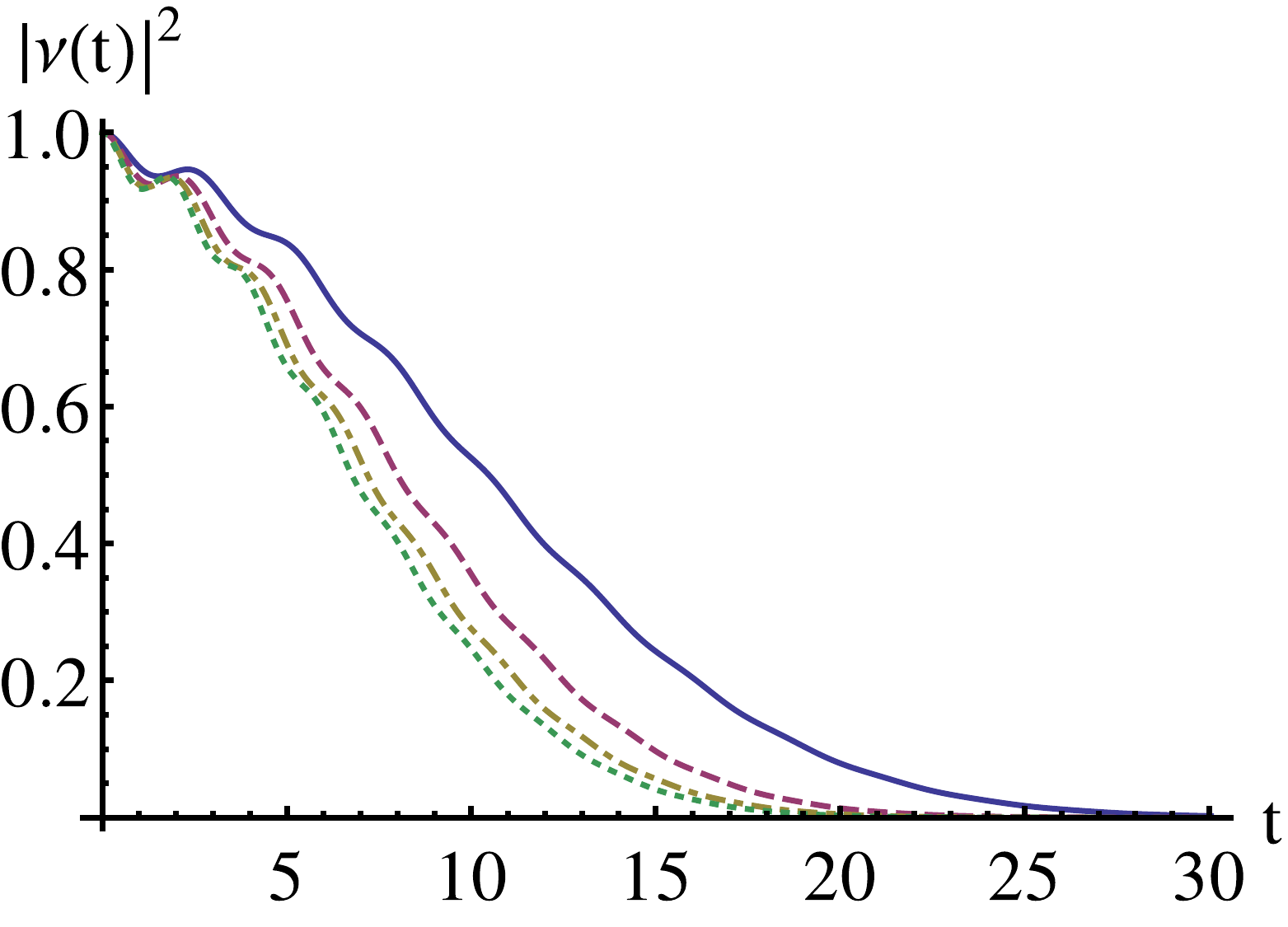}
\caption{(Color online) ${\cal L}(t)$, after a 
quench from $\lambda_0=0.5$ to $\lambda=0.55$, {for different values of $U=U_0$}, (top) {for small values}: $U=0.01$ (solid line), $U=0.02$ (dashed line),
$U=0.03$ (dot-dashed line), $U=0.04$ (dotted line), 
and (bottom) {for larger values}: $U=0.1$ (solid line), $U=0.2$ (dashed line), $U=0.3$ (dot-dashed line), $U=0.4$ (dotted line).}
 \label{fig:echoquenchlam}
\end{figure}

\subsection{Quenches in $\lambda$}
When quenching in $\lambda$ the situation is
quite different.
As we can see from Fig. \ref{fig:echoquenchlam}
the echo decays to zero, with an almost gaussian tail, 
in a finite time
and the characteristic time decay is set by $U$, namely
the larger is $U$, the faster is the decay of the echo.
This means that a quench in $\lambda$ has the effect of
making the system to explore a very large portion of
the accessible phase space, contrary to the case of the quench in $U$ where
the system remains trapped in a smaller region of the same space.

\section{Conclusions}
In this paper we report on the study of quantum quenches in
a system of ultracold bosons in a bichromatic optical lattice, 
described by a Bose-Hubbard model, in
the Bogoliubov approximation.
In particular we looked at the dynamics of the density-density
correlation functions at different times and at the Loschmidt
echo following a quench in the on-site boson-boson interaction $U$
or in the strength of the optical lattice $\lambda$.
We found that when quenching in $U$ at low $\lambda$ 
the spreading of correlation functions is ballistic 
with a speed, which is independent of $\lambda$.
By increasing $\lambda$ the signal becomes more noisy
due to the fragmentation of the energy spectrum.
{Moreover, as shown by the Loschimdt echo, after a quench in $U$, 
the final state has a large overlap with the initial one, which, unexpectedly, 
can be even larger increasing $\lambda$.}\\
On the other hand, when quenching in $\lambda$ 
at different $U$ the spreading of correlations goes from a disordered to 
a ballistic motion as $U$ increases. 
Moreover, the system seems to end in a state, which is completely
orthogonal to the initial one as witnessed by the echo.

\acknowledgments
We acknowledge financial support from MIUR through FIRB Project RBFR12NLNA\_02 and PRIN Project 2010LLKJBX.


\end{document}